\documentclass[journal]{IEEEtran}

\ifCLASSINFOpdf
\else
   \usepackage[dvips]{graphicx}
\fi
\usepackage{url}

\usepackage{mathrsfs}
\usepackage{graphicx}
\usepackage{cite}
\usepackage{amssymb}
\usepackage{amsmath}
\usepackage{booktabs}
\usepackage{makecell}
\usepackage{multirow}
\usepackage{pifont} 
\newcommand{\cmark}{\ding{51}}
\newcommand{\xmark}{\ding{55}}

\usepackage{hyperref}  
\hypersetup{
	colorlinks=false,   
	linkcolor=red,    
	urlcolor=red      
}

\usepackage[table]{xcolor}

\definecolor{ANNRow}{RGB}{245,248,255}  
\definecolor{BNNRow}{RGB}{255,248,238}  

\renewcommand{\baselinestretch}{0.93}

\begin{document}

\title{Back to Mel: Learning to Bridge Fidelity and Perception in Generative Speech Enhancement}

\author{Andong Li,~\IEEEmembership{Member, IEEE}, Kun Zhou,~\IEEEmembership{Member, IEEE}, Xiaodong Li, Chengshi Zheng,~\IEEEmembership{Senior Member, IEEE}
\thanks{Andong Li, Xiaodong Li, Chengshi Zheng are with the Institute of Acoustics, Chinese Academy of Sciences, Beijing 100190, China, and are also with University of Chinese Academy of Sciences, Beijing 100190, China (e-mail:\{liandong, lxd, cszheng\}@mail.ioa.ac.cn). Kun Zhou is with the Laboratoire de Traitement et Communication de I’Information (LTCI), Télécom Paris, Institut Polytechnique de Paris, 91120 Palaiseau, France (e-mail: kun.zhou@telecom-paris.fr). Chengshi Zheng is the corresponding author. This work was supported by the National Natural Science Foundation of China (NSFC) under Grants 62501588 and Young Elite Scientists Sponsorship Program of the Beijing High Innovation Plan. The work done by Kun Zhou was benefited from State funding managed by the French National Research Agency (ANR) under the France 2030 program, reference ANR-23-IACL-0005.}
}

\maketitle

\begin{abstract}
Recent generative techniques have facilitated the rapid proliferation of speech enhancement approaches built upon learnable compact latent representations and large generative backbones. Nevertheless, it remains inherently challenging to balance between acoustic fidelity, perceptual quality, and processing efficiency. In this paper, we revisit the Mel spectrum as a structured acoustic representation and propose BAME, a flow-matching-based framework that exploits the complementary priors for generative speech enhancement task. First, leveraging the explicit hierarchical frequency distribution of Mel spectrum, we elaborately devise a subband-aware Transformer (SaT) for efficient generative modeling. Using only 2.9\% of the parameters, the proposed SaT achieves superior enhancement performance compared with a 331 M DiT backbone. Second, motivated by range-null decomposition theory of Mel filtering in the magnitude-spectral domain, we introduce a phase-aware neural vocoder that incorporates degraded-phase prior, together with a quality-guided dual-phase strategy to adaptively mitigate the effects of unreliable phase information. Third, we devise a simple yet effective one-step fidelity-perception alignment strategy that fine-tunes the flow model with complementary constraints from fidelity and perception domains. Extensive experiments are conducted on DNS-Challenge, EARS, GSR, and URGENT benchmarks, and both objective and subjective results demonstrate promising performance on various intrusive and non-intrusive evaluation metrics. These results further advocate the potential of exploiting structured acoustic representations for reliable and efficient generative speech enhancement. Audio samples and codes are available at https://github.com/Andong-Li-speech/BAME.
\end{abstract}


\IEEEpeerreviewmaketitle
\section{Introduction}
Speech enhancement (SE) aims to recover target speech from various acoustic degradations, such as environmental noise, room reverberation and bandwidth loss~{\cite{zheng2023sixty,zhang24h_interspeech}}. Over the past decade, advances in deep neural networks (DNNs) have driven a paradigm shift from conventional statistical approaches to data-driven methods, substantially improving speech quality and intelligibility~{\cite{fan2026dbhn,li2023general,rong2024gtcrn}}. Nevertheless, most DNN-based methods follow a discriminative paradigm, learning  ``many-to-one'' deterministic mappings from degraded speech to its clean counterpart, which often struggle to reconstruct fine-grained acoustic details~{\cite{richter2023speech}}. 

Generative models have become the research hotspot in this field for their ability to model the distribution of target speech. Early studies primarily relied on generative adversarial networks (GANs), which can suffer from training instability and mode collapse~{\cite{pascual17_interspeech,shao2025cleanmel}}. More recently, large language models (LLMs) and diffusion-based models have emerged as prominent paradigms for generative speech enhancement. LLM-based approaches typically predict clean speech tokens from discrete representations obtained through neural audio codecs (NACs), followed by waveform reconstruction using codec decoders~{\cite{li24ea_interspeech,yang24h_interspeech,mu2025continuous}}. Despite their promising perceptual quality, discrete tokenization may discard fine-grained acoustic information and compromise reconstruction fidelity~{\cite{li2024autoregressive,kammoun2026modeling,fu2026rethinking}}. Meanwhile, diffusion-based generative models enable the direct modeling of continuous speech representations. Early research lines employed denoising diffusion probabilistic models (DDPMs)~{\cite{ho2020denoising,lu2022conditional}} and score-based generative models (SGMs)~{\cite{song2021score,richter2023speech,lemercier2023storm}}, while more recent flow-matching (FM) and bridge-based models offer potentially more efficient sampling through suitable probability paths and direct modeling between source and clean speech distributions~{\cite{wang25s_interspeech,wang2026flowsegrpo,jukic24_interspeech,li2025bridgevoc}}. 

Beyond generative modeling paradigms, acoustic representations play a crucial role in speech enhancement. Existing approaches can be broadly categorized into three classes. Raw representations, including waveforms and complex spectra, theoretically allow perfect reconstruction but may pose modeling challenges due to their high dimensionality and redundancy~\cite{lu2022conditional,richter2023speech}. Handcrafted compressed representations, \emph{e.g.}, Mel spectrum, reduce spectral dimension through perceptually motivated filterbanks and have been widely adopted for speech enhancement with neural vocoders~\cite{shao2025cleanmel,wang25s_interspeech}. However, their relatively high feature dimension and lack of explicit phase information may limit modeling efficiency and reconstruction fidelity, respectively. More recently, learned latent representations, such as variational autoencoders (VAEs)~\cite{zhang2026voicebridge} and neural audio codecs (NACs)~\cite{mu2025continuous}, have attracted increasing attention. By learning compact representations that retain salient speech information, these approaches can facilitate generative modeling while leveraging powerful neural decoders to produce perceptually natural speech.

Despite these advances, \textbf{learned latent representations are not necessarily optimal for speech enhancement}. Many were originally designed for audio compression~{\cite{defossez2023high}} or speech synthesis~{\cite{zhang2024speechtokenizer}}, where preserving fine-grained acoustic details is not always the primary objective. Recent studies have incorporated denoised self-supervised learning (SSL) representations to improve linguistic consistency and mitigate speech hallucinations~{\cite{rong2026pase}}. However, semantic fidelity alone does not guarantee acoustic fidelity, which also requires preserving speaker characteristics, prosodic variations, and fine-grained spectral details. Therefore, there is a significant need to reconsider acoustic representations for generative speech enhancement, with a particular emphasis on balancing reconstruction fidelity and perceptual quality. In this paper, we revisit Mel spectrum from a representation-centric perspective, aiming to exploit the under-utilized acoustic properties for high-fidelity generative speech enhancement.

\begin{figure*}[t]
	\centering
	\includegraphics[width=0.84\textwidth]{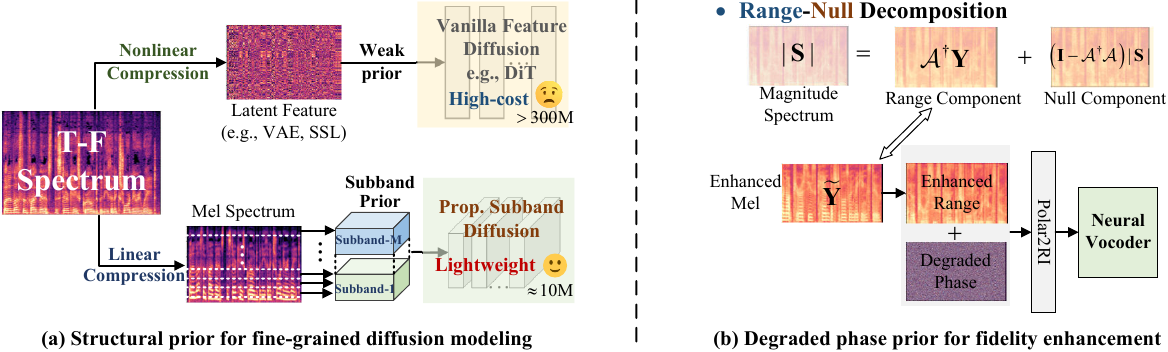}
	\caption{Illustration of two complementary priors exploited in BaME. (a) Structural prior: the explicit hierarchical frequency distribution of Mel representations facilitates fine-grained subband modeling with a lightweight generative backbone. (b) Degraded phase prior: range-null decomposition enables the recovery of magnitude information from Mel representation, which can be combined with degraded phase prior for high-fidelity waveform reconstruction.
	}
	\label{fig:system_diagram}
	\vspace{-0.40cm}
\end{figure*}

\textit{Why revisit Mel?} The rationale is twofold. First, Mel spectrum offers a \textbf{structural prior} that combines signal-related acoustic structures with compact representations. As illustrated in Fig.~{\ref{fig:system_diagram}}(a), Mel extraction can be formulated as a structured linear compression of the linear-frequency magnitude spectrum{\footnote{The logarithmic operation can be ignored herein as it is invertible.}}. This preserves an explicit hierarchical frequency distribution. Second, Mel representations enable the effective incorporation of a degraded \textbf{phase prior} for high-fidelity reconstruction. With range-null decomposition (RND) theory~{\cite{li2025learning}}, Mel spectrum can be converted into the range-space component of the linear-scale magnitude spectrum via a pseudo-inverse projection. As shown in Fig.~{\ref{fig:system_diagram}}(b), this allows the recovered magnitude information to be combined with phase prior from degraded observations, providing additional constraints for the neural vocoder to improve enhancement fidelity.

Motivated by these insights, we propose a flow-matching-based generative SE framework dubbed \textbf{BaME} (short for \underline{\textbf{Ba}}ck to \underline{\textbf{ME}}l), which explicitly exploits the complementary priors brought by Mel representations. Following a typical Mel enhancement and vocoding pipeline, BaME incorporates three key designs: \textcircled{1} We develop a subband-aware Transformer (SaT) as the backbone that exploits the explicit frequency-band structure of Mel spectrum for fine-grained and lightweight generative modeling. Despite using only \textbf{2.9\%} of the parameters and \textbf{66.6\%} of the computational complexity of DiT-331M, the proposed backbone achieves superior enhancement performance. \textcircled{2} We propose a phase-aware neural vocoder that incorporates degraded phase information into Mel-conditioned waveform reconstruction. \textcircled{3} We develop a simple yet effective one-step fidelity-perception alignment (OS-FPA) strategy that enables efficient speech generation with a single step. By jointly imposing fidelity- and perception-domain constraints, the proposed strategy effectively aligns both acoustic fidelity and perceptual quality. Extensive experiments on the DNS-Challenge, EARS, GSR, and URGENT benchmarks demonstrate that BaME consistently outperforms competitive baselines across a wide range of intrusive and non-intrusive metrics, achieving state-of-the-art performance. 

The remainder of this paper is organized as follows. In Sec.~{II}, we review related work. We present the proposed method in Sec.~III. Experimental setups are given in Sec.~{IV}. Experimental results and analysis are shown in Sec.~V, and we draw some conclusions in Sec.~VI.
\vspace{-0.25cm}
\section{Related Work}\label{sec:related-work}
In this section, we will briefly review related research work on generative speech enhancement and one-step methods.
\vspace{-0.25cm}
\subsection{Generative Speech Enhancement}\label{sec:generative-se}
Existing generative SE methods can be broadly categorized into discrete and continuous methods according to their target representations. Discrete methods typically adopt NACs to quantize clean speech into acoustic tokens for conditional generation. MaskSR~{\cite{li24ea_interspeech}} adopted masked generative modeling (MGM) to iteratively predict clean acoustic tokens, while its extended version MaskSR2~{\cite{liu2025joint}} further incorporated semantic knowledge distillation to improve speech intelligibility. AnyEnhance~{\cite{zhang2025anyenhance}} extended MGM with prompt guidance and self-critic for diverse speech and singing voice restoration tasks. Alternatively, autoregressive (AR) language models have been widely adopted for token prediction. GenSE~{\cite{yao2025gense}} adopted hierarchical modeling to sequentially generate semantic and acoustic tokens. LLaSE-G1~{\cite{kang2025llase}} predicted acoustic tokens conditioned on continuous WavLM representations.

Continuous methods, in contrast, operate on continuous speech representations. Early approaches such as VoiceFixer~{\cite{liu2022voicefixer}} adopted a two-stage paradigm that combines discriminative Mel estimation with a neural vocoder. Subsequently, CDiffuSE~{\cite{lu2022conditional}} and SGMSE~{\cite{richter2023speech}} introduced diffusion modeling in the waveform and complex spectrum domains, respectively. More recently, with the proliferation of flow-matching, FlowSE~{\cite{wang25s_interspeech}} employed FM for Mel spectrum generation, while SenSE~{\cite{li2026sense}} further combined language-model-based semantic guidance to improve the semantic consistency. Besides, bridge models have also gained increasing attention. For example, Zhang \emph{et al.}~{\cite{zhang2026voicebridge}} employed a Schrodinger bridge to establish the mapping trajectory between degraded and target latent features. To mitigate speech hallucinations, recent studies have leveraged SSL representations. For example, PASE~{\cite{rong2026pase}} leveraged denoised WavLM representations for linguistically faithful reconstruction, while subsequent extensions incorporated FM-based generative modeling to further improve perceptual quality~{\cite{rong2026stupase,gao26e_interspeech}}. 
\vspace{-0.25cm}
\subsection{One-step Generative Speech Enhancement}\label{sec:one-step}
Recently, one-step diffusion generative models have attracted increasing attention due to their notable advantage in generation efficiency. Early approaches primarily relied on distilling pre-trained teacher models into one-step generators~{\cite{salimans2022progressive}}, while consistency models enabled efficient generation by enforcing trajectory consistency~{\cite{song2023consistency}}. More recently, distillation-free approaches, such as shortcut models~{\cite{frans2025one}} and MeanFlow~{\cite{geng2025meanflow}}, have emerged to directly learn long-range transitions for one-step generation. 

One-step generation methods have recently been introduced into SE task. SE-Bridge~{\cite{qiu2023sebridge}} combined consistency modeling with Brownian bridge to enable one-step restoration, while Thunder~{\cite{trachu2024thunder}} unified regression and Brownian bridge with a single reverse step. More recently, MeanFlow has been explored in~{\cite{li2026mean}}. Alternatively, DriftSE~{\cite{xu2026driftse}} formulated SE as a generative drifting process, enabling one-step inference through distribution alignment. Despite these advances, it remains challenging to achieve high reconstruction fidelity and perceptual quality under a strict one-step inference setup.
\vspace{-0.4cm}
\section{Methodology}\label{sec:method}
\subsection{Mel Formulation}\label{sec:mel-formulation}
Despite recent advances in generative speech enhancement, the choice of acoustic representations remains largely empirical, with their intrinsic properties often under-explored in model design. In this paper, we revisit the mathematical formulation of Mel spectrum. 

Formally, given the complex-valued spectrum $\mathbf{S}\in\mathbb{C}^{F\times L}$ in the time-frequency (T-F) domain, where $F$ and $L$ denote the number of frequency bins and frames, respectively, log-scale Mel spectrum can be calculated as $\mathbf{X} = \log\left(\mathcal{A}\left|\mathbf{S}\right|\right)\in\mathbb{R}^{F_{m}\times L}$, where $\mathcal{A}\in\mathbb{R}^{F_{m}\times F}$ denotes the Mel filter, and $F_{m}$ is the number of Mel bands and usually satisfies $F_{m}\ll F$, $\left|\cdot\right|$ denotes the modulus operation. Removing the logarithmic operation, we obtain:
\begin{equation}
	\label{eqn:1}
	\mathbf{Y} = \exp\left(\mathbf{X}\right) = \mathcal{A}\left|\mathbf{S}\right|.
\end{equation} 

The Mel filterbank is constructed on a perceptually motivated frequency scale, with finer frequency resolution at lower frequencies and progressively coarser resolution at higher frequencies. Specifically, the $m$-th Mel band can be expressed as a weighted aggregation of local spectral magnitudes:
\begin{equation}
	\label{eqn:2}
	Y_{m,l} = \sum_{f\in\mathcal{B}m}\mathcal{A}_{m,f}|S_{f,l}|,
\end{equation}
where $\mathcal{B}_m$ denotes the set of frequency bins covered by the $m$-th Mel filter. Consequently, Mel compression retains the explicit frequency-band organization of the original magnitude spectrum, while progressively aggregating fine-grained spectral details toward higher frequencies, as illustrated in Fig.~{\ref{fig:system_diagram}}(a). Given the prevalence of subband modeling in the T-F domain-based speech processing tasks~{\cite{yu2022dbtnet,luo2023music}}, we thus argue that \textbf{Mel spectrum can be regarded as a coarse counterpart of T-F spectrum and modeled in a similar subband manner~{\cite{shao2025cleanmel}}}.

On the other hand, Eq.~{(\ref{eqn:1})} reveals that Mel extraction constitutes an under-determined linear mapping from the magnitude spectrum to its compressed observation. Based on range-null decomposition (RND) theory~{\cite{li2025learning}}, the magnitude spectrum can be decomposed as:
\begin{equation}
	\left|\mathbf{S}\right| = \mathcal{A}^{\dagger}\mathbf{Y}
	+
	(\mathbf{I}-\mathcal{A}^{\dagger}\mathcal{A})|\mathbf{S}|,
\end{equation} 
where $\mathcal{A}^{\dagger}$ denotes the Moore-Penrose pseudoinverse of $\mathcal{A}$. The two terms correspond to the range-space (RS) and null-space (NS) components, respectively. The RS component can be directly determined by the Mel spectrum, whereas the NS component contains the missing spectral information that can be recovered by neural vocoder, as shown in Fig.~{\ref{fig:system_diagram}}(b). Notably, although Mel spectra discard explicit phase information, degraded speech provides complementary phase cues that are often neglected in conventional Mel-based vocoding. Concretely, given the enhanced Mel spectrum $\tilde{\mathbf{Y}}$ estimated from degraded speech $\mathbf{Z}$, we construct a phase-guided complex spectral estimate
\begin{equation}
	\label{eqn:4}
	\tilde{\mathbf{S}}^{\left(0\right)} = \left|\mathcal{A}^{\dagger}\tilde{\mathbf{Y}}\right|e^{j\mathbf{\Phi}_{Z}}\in\mathbb{C}^{F\times L}
\end{equation}
as the input to a subsequent pretrained neural vocoder, where $\mathbf{\Phi}_{Z}$ denotes the phase from degraded speech $\boldsymbol{z}$.
\vspace{-0.25cm}
\subsection{Flow Matching Preliminaries}\label{sec:fm}
Flow matching (FM)~{\cite{lipman2023flow}} learns a continuous transport between a simple source distribution and the target data distribution by estimating a time-dependent velocity field. Specifically, let $\boldsymbol{\epsilon}\sim\mathcal{N}\left(\mathbf{0}, \mathbf{I}\right)$ denote Gaussian noise and $\boldsymbol{x}_{1}$ denote the target data. Following rectified flow~{\cite{liu2023flow}}, the interpolation path is defined as:
\begin{equation}
	\label{eqn:5}
	\boldsymbol{x}_{t} = \left(1 - t\right)\boldsymbol{\epsilon} + t\boldsymbol{x}_{1},\quad
	\boldsymbol{u}_{t} = \boldsymbol{x}_{1} - \boldsymbol{\epsilon},
\end{equation}
where $t\in\left[0, 1\right]$ and $\boldsymbol{u}_{t}=d\boldsymbol{x}_{t}/dt$ denotes the target velocity. Conventionally, FM trains a velocity network $\boldsymbol{v}_{\theta}$ by minimizing 
\begin{equation}
	\label{eqn:6}
	\mathcal{L}_{\mathrm{FM}}^{\left(vel\right)} = \mathbb{E}_{t,\boldsymbol{\epsilon},\boldsymbol{x}_{1}}\left[\left\|\boldsymbol{v}_{\theta}(\boldsymbol{x}_t,t,\boldsymbol{c})-\boldsymbol{u}_t\right\|_F^2\right],
\end{equation}
where $\boldsymbol{c}$ denotes the conditioning information extracted from degraded speech. Recently, Li \emph{et al.}~{\cite{li2026back}} have advocated directly predicting target data instead of velocity. Specifically, given the data prediction $\tilde{\boldsymbol{x}}_{1,\theta}=\mathcal{F}_{\theta}\left(\boldsymbol{x}_{t},t,\boldsymbol{c}\right)$, where $\mathcal{F}_{\theta}\left(\cdot\right)$ indicates the mapping instantiated by a neural network, the corresponding velocity can be derived as:
\begin{equation}
	\label{eqn:7}
	\tilde{\boldsymbol{v}}_{\theta}(\boldsymbol{x}_t,t,\boldsymbol{c})
	=\frac{\tilde{\boldsymbol{x}}_{1,\theta}-\boldsymbol{x}_t}{\max\left(1-t, \tau\right)},
\end{equation}
where $\tau$ is a small value (empirically set to 0.05) to prevent zero division. Accordingly, the velocity-matching objective can be rewritten as:
\begin{equation}
	\label{eqn:8}
	\mathcal{L}_{\mathrm{FM}}^{\left(data\right)}
	=\mathbb{E}_{t,\boldsymbol{x}_{0},\boldsymbol{x}_{1}}\left[
	\frac{\left\|\tilde{\boldsymbol{x}}_{1,\theta}-\boldsymbol{x}_1\right\|_F^2}{(1-t)^2}
	\right],
\end{equation}
which corresponds to a time-weighted data prediction loss. During inference, samples are generated by integrating the learned velocity field from $t=0$ to $t=1$. Taking discrete Euler solver as an example, the update procedure can be given by:
\begin{equation}
	\label{eqn:9}
	\tilde{\boldsymbol{x}}_{t_{\left(k+1\right)}} = \tilde{\boldsymbol{x}}_{t_{k}} + \left(t_{k+1} - t_{k}\right)\tilde{\boldsymbol{v}}_{\theta}\left(\boldsymbol{x}_{t_{k}}, t_{k}, \boldsymbol{c}\right).
\end{equation}
\begin{figure*}[t]
	\centering
	\includegraphics[width=0.78\textwidth]{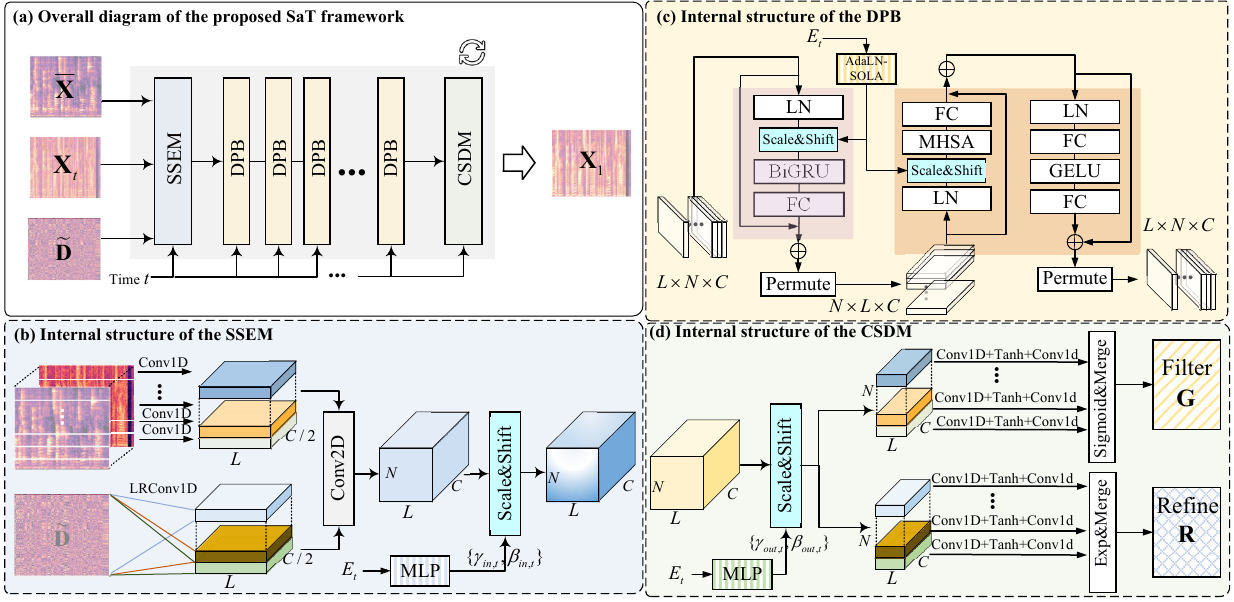}
	\vspace{-0.15cm}
	\caption{Illustration of the proposed framework for Mel enhancement. (a) Overall diagram of the proposed SaT diffusion network. Degraded Mel spectrum $\overline{\mathbf{X}}$ and $\tilde{\mathbf{D}}$ are acoustic and semantic conditions, respectively, and $\mathbf{X}_{t}$ denotes the interpolated state at timestep $t$. The network consists of synergistic subband encoding module (SSEM), multiple dual-path blocks (DPBs) and collaborative subband decoding module (CSDM). (b) Internal structure of the SSEM. (c) Internal structure of the DPB. (d) Internal structure of the CSDM. Different modules are indicated with different colors for better illustrations.
	}
	\label{fig:proposed-framework}
	\vspace{-0.40cm}
\end{figure*}
\vspace{-0.85cm}
\subsection{Proposed SaT Network}\label{sec:proposed-diffusion}
In Fig.~{\ref{fig:proposed-framework}}(a), we present the proposed subband-aware transformer (SaT) network, which aims to achieve fine-grained diffusion modeling with lightweight parameters and computational cost. It consists of three parts: synergistic subband encoding module (SSEM), stacked dual-path blocks (DPBs) and collaborative subband decoding module (CSDM). Detailed illustrations are given below.
\subsubsection{SSEM} Despite the promising perceptual quality of generative SE methods, they may suffer from inherent speech hallucinations under severe acoustic degradations, such as heavy noise and reverberation. Recently, Rong \emph{et al.} proposed to incorporate the phonological prior of WavLM to alleviate the hallucination~{\cite{rong2026pase}}. Concretely, they fine-tune a pretrained WavLM through denoising representation distillation, encouraging its final-layer representations extracted from degraded speech to match those of clean speech. Inspired by this, we adopt a similar strategy and extract the final-layer semantic representations of the adapted WavLM~{\cite{chen2022wavlm}} as auxiliary features, denoted by $\tilde{\mathbf{D}}\in\mathbb{R}^{L\times C_{D}}$, where $C_{D}$ is the channel dimension of the semantic feature.

The SSEM aims to jointly encode acoustic and semantic features in the subband space, as shown in Fig.~{\ref{fig:proposed-framework}}(b). For a given timestep $t$, the interpolated noisy and original degraded log-scale Mel spectra are denoted as $\mathbf{X}_{t}$ and $\overline{\mathbf{X}}$, respectively{\footnote{To avoid ambiguity, here ``noisy'' refers to the Gaussian noise introduced during FM interpolation, whereas ``degraded'' refers to acoustic distortions arising from real-world conditions, such as noise and reverberation.}}, and $\mathbf{X}_{t}$ is given by
\begin{equation}
	\label{eqn:10}
	\mathbf{X}_{t} = \left(1 - t\right)\boldsymbol{\epsilon} + t\mathbf{X}_{1},
\end{equation}
where $\mathbf{X}_{1}$ denotes the target log-Mel spectrum. The Mel spectra are evenly split into $N$ subbands. For the $n$-th subband, acoustic features are extracted using a 1-D convolutional layer (Conv1D):
\begin{equation}
	\label{eqn:11}
	\mathbf{I}_{n}^{\left(a\right)} = \text{Conv1D}\left(\text{Cat}\left(\mathbf{X}_{t,n}, \overline{\mathbf{X}}_{n}\right)\right)\in\mathbb{R}^{L\times \frac{C}{2}},
\end{equation}  
where the subscript $n$ denotes the subband index, $\text{Cat}\left(\cdot\right)$ refers to the concatenation along the feature axis, and $C$ is the encoding channel dimension. Unlike Mel spectrum, SSL representations lack an explicit frequency-band structure. Inspired by the low-rank subband embedding strategy in~{\cite{li2026padec}}, we employ $N$ independent low-rank convolutional layers (LRConv1D) to project the semantic features into corresponding subband representations:
\begin{equation}
	\label{eqn:12}
	\mathbf{I}_{n}^{\left(s\right)} = \text{LRConv1D}\left(\tilde{\mathbf{D}}\right)\in\mathbb{R}^{L\times \frac{C}{2}},
\end{equation}
where each LRConv1D comprises two consecutive Conv1D layers with an intermediate bottleneck to reduce parameter overhead. Subsequently, the acoustic and semantic features are fused through a point-wise 2-D convolutional layer (Conv2D) and concatenated into 3-D representation $\mathbf{H}_{in}${\footnote{The window shift size of the Mel spectrum is specially set to guarantee the same frame rate as the semantic feature.}}:
\begin{equation}
	\label{eqn:13}
	\mathbf{H}_{n}=\text{Conv2D}\left(
	\text{Cat}(\mathbf{I}_{n}^{\left(a\right)},\mathbf{I}_{n}^{\left(s\right)})
	\right)\in\mathbb{R}^{L\times C},
\end{equation}
\begin{equation}
	\label{eqn:14}
	\mathbf{H}_{in}=\text{Stack}\left(\mathbf{H}_{1},\cdots,\mathbf{H}_{N}\right)\in\mathbb{R}^{L\times N\times C},
\end{equation}
where $\text{Stack}\left(\cdot\right)$ denotes the stack operation along the subband axis. To control the encoded feature stream across different diffusion timesteps, we modulate $\mathbf{H}_{in}$ as:
\begin{equation}
	\label{eqn:15}
	\mathbf{H}_{in}^{'} = \left(1 + \gamma_{in,t}\right)\mathbf{H}_{in} + \beta_{in,t}
\end{equation} 
where $\left\{\gamma_{in,t}, \beta_{in,t}\right\}\in\mathbb{R}^{C}$ are scale and shift parameters, generated from the timestep embedding $E_{t}$.
\setcounter{figure}{3}
\begin{figure*}[t]
	\centering
	\includegraphics[width=0.90\textwidth]{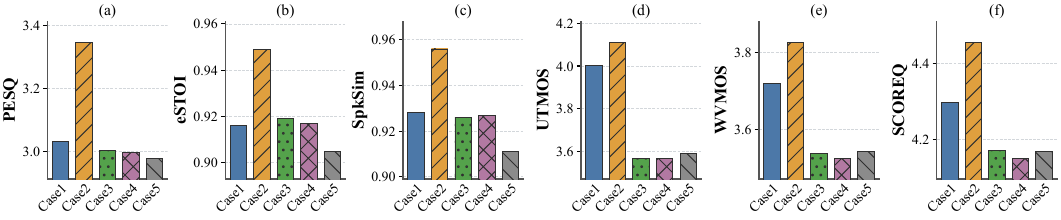}\vspace{-0.15cm}
	\caption{Performance comparisons of intrusive and non-intrusive metrics given different levels of degradation phase as auxiliary input of neural vocoder. For all cases, Mel spectra are enhanced from DNS-Challenge 2020 \textit{without reverb} test set. ``Case1'' denotes the zero-phase, ``Case2'' denotes the phase extracted from noisy speech, ``Case3'' denotes the phase extracted from reverberated speech, ``Case4'' denotes the phase extracted from the degraded speech where both environmental noise and room reverberation are involved, ``Case5'' denotes the random-phase.
	}
	\label{fig:phase-ablation}
	\vspace{-0.25cm}
\end{figure*}
\setcounter{figure}{2}
\begin{figure}[t]
	\centering
	\includegraphics[width=0.95\columnwidth]{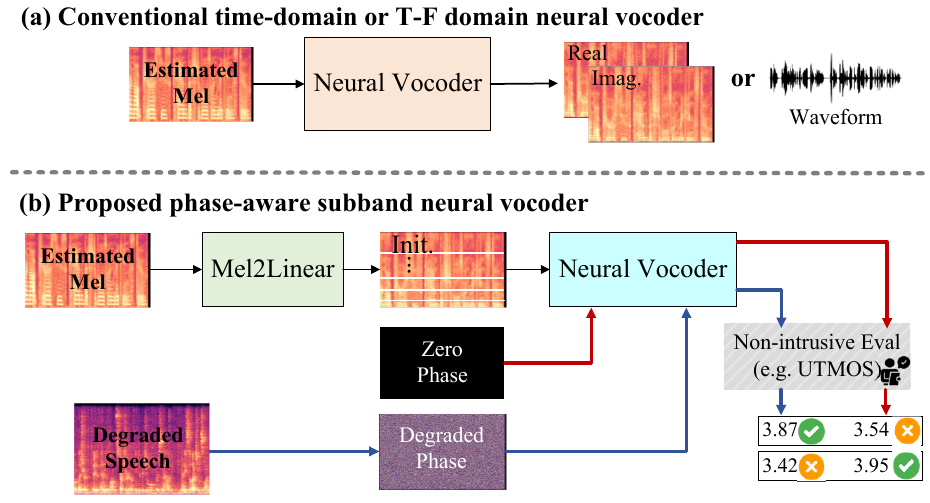}\vspace{-0.15cm}
	\caption{Generation comparisons of different neural vocoders for enhanced speech generation. (a) Conventional time-domain or T-F domain neural vocoder. (b) Proposed phase-aware subband neural vocoder.
	}
	\label{fig:phase-aware-vocoder}
	\vspace{-0.40cm}
\end{figure}
\subsubsection{DPB}\label{sec:dpb} To model the latent frame-subband relations, we stack $P$ dual-path blocks (DPBs), each of which is shown in Fig.~{\ref{fig:proposed-framework}}(c). Taking the $p$-th DPB as an example, for input feature $\mathbf{H}^{\left(p-1\right)}$, it first passes a bidirectional GRU (BiGRU) with residual connection to bridge the relations between subbands. After that, it is permuted and processed by multi-head attention (MHSA) to bridge global time relations, followed by a feedforward network (FFN). Besides, the AdaLN-SOLA~{\cite{li2025bridgevoc}} is utilized for timestep-aware feature modulation. Note that we deliberately keep the DPB architecture simple, as the combination of GRU and MHSA provides effective contextual modeling for dual-path design. Moreover, both are well-established and widely supported operations, facilitating practical implementation and deployment. In Sec.~{\ref{sec:effect-backbone}}, we further compare between different module combinations.
\subsubsection{CSDM} In contrast to SSEM, CSDM attempts to decode the target log-Mel spectrum from the subband representation, as shown in Fig.~{\ref{fig:proposed-framework}}(d). Inspired by the collaborative processing mechanism in speech enhancement~{\cite{li2022glance}}, we devise a ``\textit{filtering-and-refinement}'' decoding strategy. Specifically, given the input $\mathbf{O}\in\mathbb{R}^{\times{L\times N\times C}}$, it is first modulated by $\left\{\gamma_{out}, \beta_{out}\right\}$, and two output branches are utilized. Taking the filtering branch as an example, each subband feature is processed via a Conv1D, Tanh activation and another Conv1D layer. After that, the sigmoid function is adopted to estimate the gain coefficient $\mathbf{G}\in\mathbb{R}^{F_{m}\times L}$. For the refinement branch, the similar operation is conducted to estimate the refinement component $\mathbf{R}\in\mathbb{R}^{F_{m}\times L}$ except the exponential function is utilized. Then, the estimated log-Mel spectrum is obtained as:
\begin{equation}
	\label{eqn:16}
	\tilde{\mathbf{X}}_{1,\theta} = \log\left(\exp(\overline{\mathbf{X}})\odot\mathbf{G} + \mathbf{R}\right),
\end{equation}
where $\odot$ denotes element-wise multiplication operation. Note that we directly predict the data and adopt the loss in Eq.~{(\ref{eqn:8})}.

\textbf{Remarks}: Although recent work CleanMel~{\cite{shao2025cleanmel}} also investigates Mel-domain subband modeling, the proposed SaT differs from it in three key aspects. First, CleanMel performs narrow-band modeling at individual Mel bins, whereas the proposed SaT groups adjacent Mel bands into a single subband, substantially reducing computational overhead. In Sec.~{\ref{sec:effect-backbone}}, we reveal that excessively fine-grained subband partitioning may provide limited performance gains, suggesting modeling redundancy among adjacent Mel bands. Second, while CleanMel adopts a discriminative Mel estimation paradigm, \textbf{our SaT is the first Mel subband-aware backbone specially designed for generative SE tasks}. Third, CleanMel employs Mamba for temporal modeling, whose commonly used optimized implementation relies on specialized CUDA kernels. In contrast, SaT adopts conventional GRU and MHSA operations, offering broader compatibility with standard deep learning runtimes and facilitating deployment on CPU-oriented platforms.
\setcounter{figure}{4}
\begin{figure*}[t]
	\centering
	\includegraphics[width=0.85\textwidth]{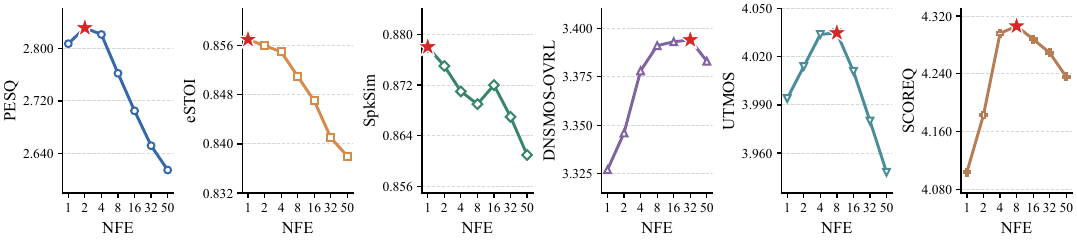}\vspace{-0.15cm}
	\caption{Metric performance under different NFEs. Scores are calculated and averaged upon the DNS-Challenge 2020 \textit{without reverb} and \textit{with reverb} test sets.
	}
	\label{fig:metric-nfe}
	\vspace{-0.40cm}
\end{figure*}
\begin{figure*}[t]
	\centering
	\includegraphics[width=0.70\textwidth]{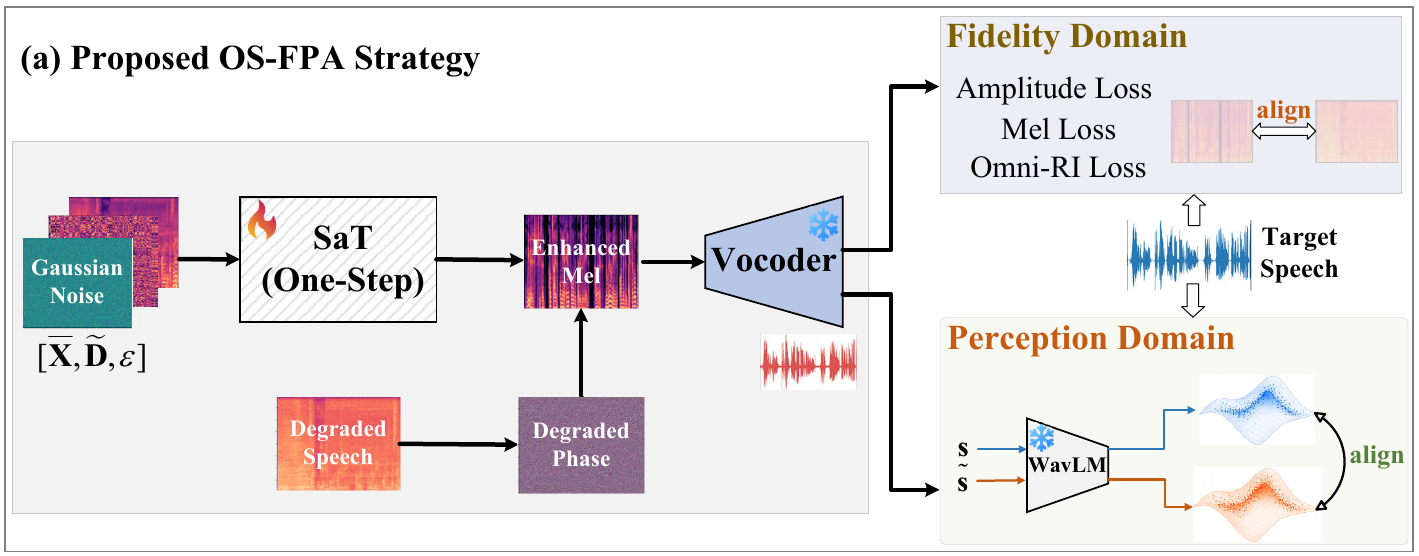}
	\vspace{-0.15cm}
	\caption{Diagram of the proposed OS-FPA scheme. Gaussian noise and conditioning feature serves as input to SaT to obtain the enhanced Mel spectrum with a single step, and the enhanced waveform is generated using a pretrained neural vocoder. Two domains are employed to encourage fidelity preservation and perceptual quality, respectively. For the former, amplitude loss, mel-loss and omnidirectional RI-loss are adopted. For the latter, both enhanced and target signals are passed to the pretrained WavLM and the latent features are forced to align.
	}
	\label{fig:osfpa}
	\vspace{-0.40cm}
\end{figure*}
\vspace{-0.35cm}
\subsection{Phase-aware Neural Vocoder}\label{sec:phase-neural-vocoder}
Existing Mel-based generative SE methods typically reconstruct enhanced speech from the estimated Mel spectrum using a pre-trained neural vocoder, as illustrated in Fig.~{\ref{fig:phase-aware-vocoder}}(a). However, Mel representations discard explicit phase information, requiring the vocoder to infer the missing phase from the learned speech prior. Previous studies have demonstrated the significance of phase information for speech enhancement~{\cite{oppenheim1981importance,gerkmann2015phase}}. Unlike speech synthesis, SE provides degraded observations as an auxiliary prior, and simply neglecting it may compromise reconstruction fidelity.

Recently, we developed subband-based neural vocoders called DegVoC~{\cite{li2026degvoc}}, which exploit the degradation structure of Mel representations for efficient waveform generation. Specially, the Mel spectrum is first mapped to the linear-scale T-F domain as the RS component to obtain an initial magnitude estimate, which is then combined with zero-phase initialization to construct a complex spectrum. A lightweight network then performs subband encoding, contextual modeling and spectral decoding to recover the target spectrum. This design enables fine-grained spectral reconstruction with relatively low computational complexity. To further exploit the prior in SE, we extend DegVoC by incorporating degraded phase into the spectral initialization, as illustrated in Fig.~{\ref{fig:phase-aware-vocoder}}(b). Specifically, given the enhanced log-scale Mel spectrum $\mathbf{\tilde{X}}_{1}$, the reconstructed waveform $\tilde{\boldsymbol{s}}$ can be given by:
\begin{equation}
	\label{eqn:17}
	\tilde{\boldsymbol{s}}=\mathcal{V}_{\psi}\left(\tilde{\mathbf{M}}\odot\cos\left(\mathbf{\Phi}_{Z}\right), \tilde{\mathbf{M}}\odot\sin\left(\mathbf{\Phi}_{Z}\right)\right),
	\vspace{-0.25cm}
\end{equation}
where $\tilde{\mathbf{M}}=\mathcal{A}^{\dagger}\exp\left(\mathbf{\tilde{X}}_{1}\right)$ denotes the initialized magnitude spectrum, and $\mathcal{V}_{\psi}$ denotes the neural vocoder, and $\mathbf{\Phi}_{Z}$ is the degraded phase. 

Nonetheless, degraded phase is not always beneficial for speech reconstruction, as it heavily relies on practical acoustic degradation scenarios. To investigate the effect of different phase initialization conditions, we train a phase-conditioned DegVoC using the clean Mel spectra and degraded phase generated through on-the-fly noise and reverberation simulation, and evaluate five phase initialization conditions: zero phase, phase extracted from noisy speech, reverberant phase, noisy-reverberant phase, and random phase. In the inference stage, the restored Mel spectra are generated by SaT with eight sampling steps (NFE=8), and five phase conditions are adopted for waveform generation.

Fig.~{\ref{fig:phase-ablation}} presents the results on six representative objective metrics, including three intrusive metrics (PESQ~{\cite{rec2005p}}, eSTOI~{\cite{jensen2016algorithm}} and SpkSim~{\cite{jung2019RawNet}}) to assess complementary aspects of reconstruction quality, intelligibility, and speaker consistency, and three non-intrusive metrics (UTMOS~{\cite{saeki2022utmos}}, WVMOS~{\cite{andreev2023hifipp}} and SCOREQ~{\cite{ragano2024scoreq}}) as proxies for perceptual quality.
Higher scores indicate better performance, and detailed illustrations of the metrics are provided in Sec.~{\ref{sec:metrics}}. Notably, noisy phase consistently yields the best performance across all metrics, indicating the potential benefit of incorporating degraded phase as an acoustic prior. However, reverberant and noisy-reverberant phase cases exhibit considerably lower non-intrusive quality scores than the zero-phase case. The phenomenon may be associated with reverberation-induced phase distortion and temporal smearing, which can introduce inconsistent acoustic cues when combined with enhanced spectral magnitudes~{\cite{luo2022analysis}}. These observations suggest that the effectiveness of degraded phase is highly constrained by acoustic conditions, which can be unstable under practical acoustic scenarios.

To tackle this problem, we propose a quality-guided dual-phase vocoding (QDV) strategy, as shown in Fig.~{\ref{fig:phase-aware-vocoder}}(b). Concretely, the enhanced Mel spectrum is decoded using both zero-phase and degraded-phase initialization, yielding two candidate waveforms $\tilde{\boldsymbol{s}}_{0}$ and $\tilde{\boldsymbol{s}}_{deg}$. A non-intrusive evaluator (NIE) is then employed to assess their perceptual quality without requiring clean reference. The final enhanced speech can be chosen as:
\begin{equation}
	\label{eqn:18}
	\begin{aligned}
		\phi^{}
		&=\arg\max_{\phi\in\{0,deg\}}
		Q_{\mathrm{NIE}}(\tilde{\boldsymbol{s}}_{\phi}),\
		\tilde{\boldsymbol{s}}
		&=\tilde{\boldsymbol{s}}_{\phi},
	\end{aligned}
\end{equation}
where $Q_{\text{NIE}}\left(\cdot\right)$ denotes the predicted quality score. By adaptively selecting between the two reconstruction candidates, the proposed strategy effectively alleviates the risk of incorporating unreliable degraded phase. Further, due to the use of lightweight DegVoC as the vocoder, the proposed scheme only introduces modest computational overhead. 
\vspace{-0.35cm}
\subsection{One-Step Fidelity-Perception Alignment Strategy}\label{sec:os-fpa}
While FM-based generative SE methods have demonstrated promising perceptual quality, they usually rely on iterative sampling, heavily impeding the processing efficiency. Moreover, we observe a notable discrepancy between intrusive and non-intrusive metrics \emph{w.r.t.} the number of function evaluations (NFEs). Specifically, we evaluate six metrics under different NFEs, including three intrusive metrics (PESQ, eSTOI, SpkSim) to assess reference-based reconstruction quality and three non-intrusive metrics (DNSMOS-OVRL~{\cite{reddy2022dnsmos}}, UTMOS and SCOREQ) to measure the perceptual quality. Higher scores indicate better performance. The proposed SaT serves as the generative network and the phase-aware neural vocoder in Sec.~{\ref{sec:phase-neural-vocoder} is employed for waveform reconstruction.

As shown in Fig.~{\ref{fig:metric-nfe}}, the three intrusive metrics achieve the highest scores at $\text{NFE}=1$ or 2, whereas larger NFEs are needed for non-intrusive metrics. These observation suggest that low-NFE generation already preserves considerable reference-based fidelity, which can benefit from the strong alignment of the degraded Mel condition. However, its predicted perceptual quality remains suboptimal and may benefit from additional generative refinement. As a result, it is significant to develop a one-step strategy that can improve both perceptual quality and reconstruction fidelity.

Motivated by~{\cite{zhao2026perceptual}}, we propose a simple yet effective one-step fidelity-perception alignment strategy called OS-FPA, whose overall diagram is shown in Fig.~{\ref{fig:osfpa}}. After training SaT with standard FM objective, we adopt fine-tuning for one-step generation. Since SaT adopts a data-prediction parameterization in Eq.~{\ref{eqn:8}}, the output of the network directly estimates the target log-scale Mel spectrum:
\begin{equation}
	\label{eqn:19}
	\tilde{\mathbf{X}}_1
	=
	\mathcal{F}_{\theta}
	\left(\boldsymbol{\epsilon},0,\boldsymbol{c}\right),
	\qquad
	\boldsymbol{\epsilon}\sim
	\mathcal{N}(\mathbf{0},\mathbf{I}),
	\vspace{-0.25cm}
\end{equation}
where $\boldsymbol{c}=\left(\overline{\mathbf{X}}, \tilde{\mathbf{D}}\right)$ denotes the condition, ``0'' denotes the beginning timestep. Following the vocoding procedure in Eq.~(\ref{eqn:17}), the predicted Mel spectrum is converted into an enhanced waveform. To accommodate both phase initialization conditions, we randomly select zero-phase or degraded-phase with equal probability during OS-FPA fine-tuning.

After that, we jointly optimize the enhanced waveform in the perception and fidelity domains. To be specific, for perceptual alignment, the enhanced speech $\tilde{\boldsymbol{s}}$ and clean speech $\boldsymbol{s}$ are passed through a frozen pretrained SSL model (\emph{e.g.}, WavLM) to extract the latent features from multiple layers. Let the features from the $k$-th Transformer layer of the WavLM be denoted by  $\tilde{\mathbf{\Theta}}^{\left(k\right)}$ and $\mathbf{\Theta}^{\left(k\right)}$, respectively, we attempt to align the corresponding feature representations to be as close as possible, that is:
\vspace{-0.25cm}
\begin{equation}
	\label{eqn:20}
	\mathcal{L}_{perc} = -\frac{1}{\left|\mathcal{K}\right|}\sum_{k\in\mathcal{K}}\frac{1}{L_{k}}\sum_{l=1}^{L_{k}}\text{CosSim}\left(\tilde{\mathbf{\Theta}}^{\left(k\right)}_{l}, \mathbf{\Theta}^{\left(k\right)}_{l}\right),
\end{equation}
where $\mathcal{K}$ denotes the selected layer set, $L_{k}$ is the number of frames, and $\text{CosSim}\left(\cdot, \cdot\right)$ calculates cosine similarity along the feature dimension. 

However, perceptual loss alone can introduce noticeable artifacts, as shown in Fig.~{\ref{fig:one-step-spec}}, indicating that SSL feature alignment may not fully constrain fine-grained spectral reconstruction. To this remedy, we introduce complementary fidelity-domain constraints, which include log-amplitude loss, multi-resolution Mel loss and Omnidirectional RI loss~{\cite{dai2026gompsnr}}, given by:
\begin{equation}
	\label{eqn:21}
	\mathcal{L}_{fid} = \lambda_{amp}\mathcal{L}_{amp} + \lambda_{mel}\mathcal{L}_{mel} + \lambda_{ori}\mathcal{L}_{ori},
\end{equation}
where $\left\{\lambda_{amp},\lambda_{mel},\lambda_{ori}\right\}$ are corresponding weighting parameters. The overall dual-domain objective can be written as:
\begin{equation}
	\label{eqn:22}
	\mathcal{L}_{OS-FPA} = \mathcal{L}_{fid} + \lambda_{perc}\mathcal{L}_{perc},
\end{equation} 
where $\lambda_{perc}$ denotes the weighting parameter of the perceptual loss.
\begin{figure}[t]
	\centering
	\includegraphics[width=0.95\columnwidth]{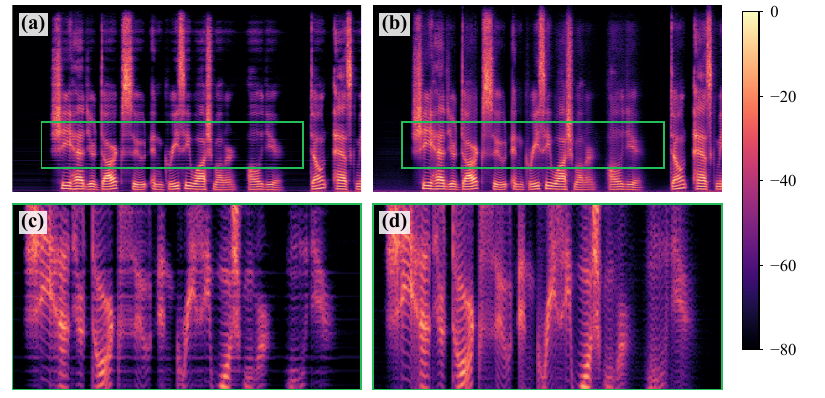}
	\vspace{-0.15cm}
	\caption{Spectral visualizations in terms of whether to include fidelity constraints. (a) Enhanced spectrum with only perceptual-domain loss. (b) Enhanced spectrum with additional fidelity-domain loss. (c)-(d) Zoom-in views of the regions highlighted by green boxes in (a)-(b), respectively. The audio file is from DNS-Challenge 2020 \textit{with reverb} test set.
	}
	\label{fig:one-step-spec}
	\vspace{-0.30cm}
\end{figure}

\begin{table}[t]
\centering
\caption{Data augmentation configurations.}
\label{tbl:degrade-augmentation}
\small
\setlength{\tabcolsep}{3pt}
\renewcommand{\arraystretch}{1.15}
\resizebox{0.8\columnwidth}{!}{
\begin{tabular}{lcc}
	\hline\hline
	Distortion & Probability & Hyperparameters \\
	\hline
	Noise &0.5 &SNR$\in[-8, 15]$ dB\\
	\hline
	Reverberation & 0.5 & -- \\
	\hline
	Wind noise & 0.08 & SNR $\in [-8, 15]$ dB \\
	\hline
	Clipping &0.25 &
	\begin{tabular}{c}
		Min\_quantile $\in [0.0,0.1]$ \\
		Max\_quantile $\in [0.9,1.0]$
	\end{tabular} \\
	\hline
	LPF &0.25 &Lower bound=1 kHz \\
	\hline
	Codec &0.25 &
	\begin{tabular}{c}
		Format $\in \{\text{mp3},\text{opus}, \text{gsm}\}$ \\
	\end{tabular} \\
	\hline
	Packet loss &0.25 &
	\begin{tabular}{c}
		Rate $\in [0.05,0.3]$ \\
		Min\_frame\_time = 8 ms \\
		Max\_frame\_time = 100 ms
	\end{tabular} \\
	\hline
	T-F Masking &0.25 &\begin{tabular}{c}
		\{Height, Width\} $\in[1, 12]$\\
		Hole number $\in[1, 150]$
	\end{tabular} \\
	\hline\hline
\end{tabular}
}
\vspace{-0.35cm}
\end{table}
\renewcommand\arraystretch{0.95}
\begin{table*}[t]
	\centering
	\Large
	\caption{Descriptions of the adopted baseline methods. ``D'' and ``G'' denote discriminative and generative SE approaches, respectively. ``-'' denotes no code is provided, and ``$\dagger$'' denotes that only a mismatched implementation is open-sourced, which is inconsistent with the original setup in the paper.}
	\label{tbl:descriptions-baselines}
	\resizebox{0.9\textwidth}{!}{
		\begin{tabular}{ccccccc}
			\hline\hline
			\textbf{Ids} &\textbf{Methods} &\textbf{Type} &\textbf{Publish} &\textbf{Representation Type} &\textbf{Code Links} &\textbf{Weights Available}\\
			\hline
			1  &TFGridNet~{\cite{wang2023tfgridnet}} &D &TASLP 2023 &Complex Spectrum &https://github.com/espnet/espnet &\cmark\\
			2 &BSRNN~{\cite{yu2023high}} &D &Interspeech 2023 &Complex Spectrum & https://github.com/espnet/espnet  &\cmark\\
			3 &StoRM~{\cite{lemercier2023storm}} &G &TASLP 2023 &Complex Spectrum &https://github.com/sp-uhh/storm &\cmark \\
			4 &VoiceFixer~{\cite{liu2022voicefixer}} &G &Interspeech 2022 &Mel &https://github.com/Render-AI-Team/voicefixer2 &\cmark \\
			5 & CleanMel~{\cite{shao2025cleanmel}} &G &TASLP 2025 &Mel &https://github.com/Audio-WestlakeU/CleanMel &\cmark \\
			6 &FlowSE~{\cite{wang25s_interspeech}} &G &Interspeech 2025 &Mel &$\dagger$& \xmark \\
			7 &SenSE~{\cite{li2026sense}} &G &ICME 2026 &Mel &https://github.com/ASLP-lab/SenSE &\cmark \\
			8 &StuPASE~{\cite{rong2026stupase}} &G &Interspeech 2026 &Mel &https://github.com/cisco-open/pase/tree/main/stupase &\cmark \\ 
			9 &SELM~{\cite{wang2024selm}} &G &ICASSP 2024 &Discrete Token  &- &\xmark\\
			10 &MaskSR~{\cite{li24ea_interspeech}} &G &Interspeech 2024 &Discrete Token &- &\xmark\\
			11 &MaskSR2~{\cite{liu2025joint}} &G &ICASSP 2025 &Discrete Token &- &\xmark \\
			12 &LLaSE-G1~{\cite{kang2025llase}} &G &ACL 2025 &Discrete Token &https://github.com/Kevin-naticl/LLaSE-G1/tree/main &\cmark \\
			13 &Metis~{\cite{wang2025metis}} &G &NeurIPS 2025 &Discrete Token &https://github.com/open-mmlab/Amphion/tree/main/models/tts/metis &\cmark \\
			14 &AnyEnhance~{\cite{zhang2025anyenhance}} &G &TASLP 2025 &Discrete Token &$\dagger$ &\xmark\\
			15 &GenSE~{\cite{yao2025gense}} &G &ICLR 2025 &Discrete Token &https://github.com/yaoxunji/gen-se &\cmark \\
			16 &UniSE~{\cite{yan2026unise}} &G &Interspeech 2026 &Discrete Token &https://github.com/alibaba/unified-audio/tree/main/QuarkAudio-UniSE &\cmark\\
			17 &L3-SE~{\cite{wang2026reducing}} &G &Arxiv 2026 &Discrete Token &- &\xmark\\
			18 &Xingchen-UniFE~{\cite{kuang2026xingchen}} &G &Arxiv 2026 &Discrete Token &- &\xmark \\
			19 &VoiceBridge~{\cite{zhang2026voicebridge}} &G &ICML 2026 &VAE &- &\xmark\\
			20 &PASE~{\cite{rong2026pase}} &G &AAAI 2026 &SSL &https://github.com/cisco-open/pase &\cmark\\
			21 &PhASE-Flow~{\cite{gao26e_interspeech}} &G &Interspeech 2026 &SSL &- &\xmark \\
			22 &UniPASE~{\cite{rong2026unipase}} &G &TASLP 2026 &SSL &https://github.com/Xiaobin-Rong/unipase &\cmark \\
			\hline\hline
	\end{tabular}}
	\vspace{-0.35cm}
\end{table*}
\vspace{-0.15cm}
\section{Experimental Setups}\label{sec:experimental-setup}
\vspace{-0.15cm}
\subsection{Datasets Setup}\label{sec:datasets}
\subsubsection{Training Datasets} We construct a large-scale training corpus. Clean speech clips are collected from multiple sources, including the LibriVox subset of the DNS5 Challenge~{\cite{dubey2024icassp}}, AISHELL3~{\cite{shi2020aishell}}, DiDi~{\cite{guo2021didispeech}}, EARS~{\cite{richter2024ears}}, VCTK~{\cite{veaux2013voicebank}}, LibriTTS~{\cite{zen2019libritts}}, MLS~{\cite{pratap2020mls}}, and four additional datasets specially provided by URGENT Challenge 2026 (ESD, NNCES, SeniorTalk and VocalSet){\footnote{https://github.com/urgent-challenge/urgent2026\_challenge\_track1.}} For the first seven datasets, we assess speech quality using UTMOS and DNSMOS-OVRL, and retain the top 25\% of clips ranked by the averaged score to improve speech quality. To prevent potential overlap with the evaluation sets, we exclude speakers p102-p107 from EARS and the last eight speakers from VCTK, following~\cite{liu2022voicefixer} for the latter. Besides, the English subset of MLS is excluded due to its relatively poor audio quality. After filtering, the total duration of the clean speech corpus is $\approx$ 600 hours. Noise clips are collected from the DNS-Challenge noise corpus, FMA~{\cite{defferrard2017fma}}, FSD50K~{\cite{fonseca2022fsd50k}} and WHAM!~{\cite{wichern2019wham}}. To prevent data overlap with EARS-WHAM\_v2 benchmark, we exclude the WHAM! test split from the training noise pool. Room impulse responses (RIRs) are sourced from openSLR26, OpenSLR28{\footnote{https://www.openslr.org/resources.php.}}, ARNI~{\cite{prawda2022calibrating}}, and BRUDEX~{\cite{fejgin2023brudex}}.

We generate degraded-target speech pairs on the fly using two degradation schemes. The first simulates additive noise and convolutive reverberation. Specifically, environmental noise and wind noise are added to each clean speech clip with probabilities of 0.5 and 0.08, respectively, with the signal-to-noise ratio (SNR) uniformly sampled from $[-8,15]$ dB. Room reverberation is applied with a probability of 0.5, with the corresponding direct-path speech serving as the target. The second scheme introduces additional acoustic degradations, including low-pass filtering (LPF), audio clipping, audio codecs, packet loss, and spectral T-F masking. For each training sample, one to three degradation operations are randomly selected and applied in a shuffled order to increase degradation diversity. Detailed configurations are summarized in Table~{\ref{tbl:degrade-augmentation}}. 
\subsubsection{Evaluation Datasets} We evaluate performance on four mainstream benchmarks, which are given below:
\begin{itemize}
	\item \textbf{DNS-Challenge 2020 test set}: It is a synthesized test set with sampling rate of 16 kHz and involves two subsets, namely \textit{without reverb} and $\textit{with reverb}$, each containing 150 degraded-clean pairs. For the former, only environmental noise is added. For the latter, both reverberation and noise are involved, corresponding to joint denoising and dereverberation task. No transcript is provided for both subsets.
	\item \textbf{EARS-WHAM\_v2} and \textbf{EARS-Reverb\_v2}: The former mixed clean speech p102-p107 from the EARS with nonstationary noise from WHAM! at SNRs ranging from -2.5 to 17.5 dB. The latter generated reverberant speech by convolving the same clean speech with RIRs from the ACE-Challenge corpus, with reverberation time (RT60) limited to 2.0 s. Each test set comprises 886 degraded-clean pairs, and are adopted to evaluate denoising and reverberation performance, respectively. No clean transcript is available. The original sampling rate of the benchmark is 48 kHz and we resample it to 16 kHz.
	\item \textbf{GSR test set}: introduced in~{\cite{liu2022voicefixer}}, the GSR test set consists of 503 degraded-clean pairs and involves four acoustic degradations: noise, reverberation, clipping and bandwidth loss. No clean transcript is available. The original sampling rate of the benchmark is 44.1 kHz and we resample it to 16 kHz.  
	\item \textbf{URGENT Challenge 2025 non-blind test set}: It is an official non-blind test set from the URGENT 2025 Challenge{\footnote{https://urgent-challenge.github.io/urgent2025.}}, which aims to evaluate the performance of unified speech enhancement (USE), \emph{i.e.}, a wide variety of acoustic distortions and different sampling rates. It includes 1000 degraded-clean pairs. All degraded and clean speech clips are resampled to 16 kHz whenever their original sampling rates differ from 16 kHz. 
\end{itemize}
\vspace{-0.45cm}
\renewcommand\arraystretch{1.00}
\subsection{Baselines}\label{sec:baselines}
We compare with various advanced baselines, covering discriminative, discrete and continuous generative SE methods. Details are summarized in Table~{\ref{tbl:descriptions-baselines}}.
\begin{table*}[t]
	\centering
	\caption{Ablation studies \emph{w.r.t.} the data-prediction loss, different modules in DPB, the adopted trajectory. ``V'' and ``D'' denote velocity-prediction and data-prediction loss, respectively. ``Traj.'' denotes the type of trajectory modeling. ``Sch-VP'' denotes the Schrodinger bridge with variance preserving scheduler, and ``Sch-VE'' denotes the Schrodinger bridge with variance exploding scheduler. All results are averaged upon DNS-Challenge 2020 \textit{without reverb} and \textit{with reverb} test sets. $\text{NFE}=8$ is adopted by default. The best and second-best results are highlighted with``\textbf{BOLD}'' and ``\textbf{UNDERLINED}'', respectively.}
	\label{tbl:ablation-arch}
	\resizebox{0.9\textwidth}{!}{
		\begin{tabular}{cccccccccccccc}
			\hline\hline
			\multirow{2}{*}{\textbf{Ids}} & \multirow{2}{*}{\textbf{Loss}} &\multirow{2}{*}{\textbf{Backbone}} &\multirow{2}{*}{\textbf{Traj.}} &\multirow{2}{*}{\textbf{N}}  & 
			\multirow{2}{*}{\textbf{PESQ}$\uparrow$} &\multirow{2}{*}{\textbf{eSTOI}$\uparrow$} &\multirow{2}{*}{\textbf{LPS}$\uparrow$} &\multirow{2}{*}{\textbf{SBS}$\uparrow$} &\multirow{2}{*}{\textbf{SpkSim}$\uparrow$} & 
			\textbf{DNSMOS}$\uparrow$ &\multirow{2}{*}{\textbf{UTMOS}$\uparrow$} & \textbf{WER}$\downarrow$ \\
			& &  & & & & & & & & \textbf{OVRL} & &\textbf{(in \%)}\\
			\hline
			1 & V & DiT  & FM &-   &2.357 &0.804 &0.883 &0.881 &0.844 &3.376 &\textbf{3.860} &13.486 \\
			2 & D & DiT  & FM  &-   &\underline{2.448} &\underline{0.815} &\underline{0.896} &\underline{0.891} &\underline{0.849} &\underline{3.390} &3.847 &11.615 \\
			\midrule
			3 & D & SaT & FM &25   &\textbf{2.487} &\textbf{0.824} &\textbf{0.907} &\textbf{0.893} &\textbf{0.854} &\textbf{3.391} &\underline{3.857} &\textbf{9.242} \\
			4 & D &Variant1 & FM &25     &2.424 &0.810 &0.891 &0.881 &0.833 &3.390 &3.797 &\underline{10.937} \\
			5 & D &Variant2 & FM &25      &2.412 &0.814 &0.893 &0.879 &0.793 &3.389 &3.773 &10.912 \\
			6 & D & Variant3 & FM &25       &2.411 &0.812 &0.894 &0.881 &0.810 &3.361 &3.782 &11.404 \\
			7 & D & Variant4 & FM  &25       &2.074 &0.754 &0.826 &0.830 &0.759 &3.362 &3.353 &18.496 \\
			\hline
			8 & D & SaT & Sch-VP &25 &2.240 &0.775 &0.870 &0.859 &0.765 &3.356 &3.407 &11.665 \\
			9 & D & SaT & Sch-VE &25    &2.293 &0.778 &0.875 &0.862 &0.761 &3.365 &3.538 &10.962 \\
			\hline\hline
		\end{tabular}
	}
	\vspace{-0.15cm}
\end{table*}
\begin{figure*}[t]
	\centering
	\includegraphics[width=0.85\textwidth]{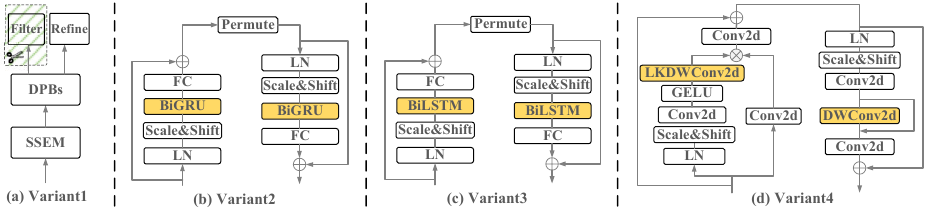}
	\vspace{-0.15cm}
	\caption{Internal structures of different backbone variants. (a) Internal structure of the variant1, where the filtering branch of the CSDM is removed. (b) Internal structure of the variant2, where the BiGRU is utilized for both frame and subband relation modeling within the DPB. (c) Internal structure of the variant3, where the BiLSTM is utilized for both frame and subband relation modeling within the DPB. (d) Internal structure of the variant4, where we replace the DPB with the convolution-based basic unit proposed in~{\cite{li2025bridgevoc}}.
	}
	\label{fig:backbone-variants}
	\vspace{-0.40cm}
\end{figure*}
\begin{table}[t]
	\centering
	\Huge
	\caption{Effect of different semantic representations. Note that the
		reported model parameters and computational complexity account for the semantic extractor, SaT with $\text{NFE=8}$, and the neural vocoder. Results are averaged upon DNS-Challenge 2020 \textit{without reverb} and \textit{with reverb} test sets.}
	\label{tbl:semantic_ablation}
	\resizebox{0.98\columnwidth}{!}{
		\begin{tabular}{cccccccc}
			\toprule
			\multirow{2}{*}{\textbf{Ids}} &\textbf{\#Params.} &\textbf{MACs} &\textbf{Semantic} &\multirow{2}{*}{\textbf{PESQ}$\uparrow$} &\multirow{2}{*}{\textbf{eSTOI}$\uparrow$} &\multirow{2}{*}{\textbf{UTMOS}$\uparrow$} &\textbf{WER}$\downarrow$ \\
			&\textbf{(M)} &\textbf{(Giga/s)} &\textbf{Extractor}  & & &  &\textbf{(in \%)}\\
			\midrule
			1 &15.45 &58.72 & \xmark &2.487 &0.824 &3.857 &9.242 \\
			2 &332.38 &73.43  & DeWavLM &\textbf{2.630} &\textbf{0.838} &\textbf{3.968} &\textbf{4.269} \\
			3 &27.14($\downarrow$91.83\%) &63.52($\downarrow$13.50\%) &SubNet &\underline{2.520} &\underline{0.830} &\underline{3.871} &\underline{8.538} \\
			\bottomrule
		\end{tabular}
	}
	\vspace{-0.66cm}
\end{table}
\vspace{-0.35cm}
\subsection{Metrics}\label{sec:metrics}
For comprehensive comparisons, we evaluate multiple intrusive and non-intrusive objective metrics, spanning speech quality, intelligibility, speaker similarity, phoneme similarity, and linguistic correctness. Here the terms \emph{intrusive} and \emph{non-intrusive} differ in whether reference is required, rather than directly corresponding to fidelity and perceptual quality. In the following parts, we use intrusive metrics as complementary indicators of reference consistency, while non-intrusive speech quality predictors serve as proxies for perceptual quality. Details are provided below.
\subsubsection{Intrusive Metrics} A time-aligned clean reference is needed to evaluate the enhanced speech.
\begin{itemize}
	\item PESQ~{\cite{rec2005p}}: Perceptual Evaluation of Speech Quality (PESQ) estimates overall speech quality. We adopt the wideband version, with score ranging from -0.5 to 4.5. 
	\item eSTOI~{\cite{jensen2016algorithm}}: Extended Short-Time Objective Intelligibility (eSTOI) measures speech intelligibility, with score ranging from 0 to 1. 
	\item LPS~{\cite{pirklbauer2023evaluation}}: Levenshtein Phoneme Similarity (LPS)
	uses a \texttt{wav2vec 2.0} model fine-tuned for phoneme recognition to evaluate phonetic similarity between reference and enhanced speech, with the range of [0, 1]. 
	\item SpkSim~{\cite{jung2019RawNet}}: Speaker similarity (SpkSim) uses a pretrained \texttt{voxcelebs12\_rawnet3}{\footnote{https://huggingface.co/espnet/voxcelebs12\_rawnet3.}} to calculate the speaker similarity between reference and enhanced speech, with the range of [0, 1].
	\item SBS~{\cite{saeki2024speechbertscore}}: SpeechBERTScore (SBS) measures the overall similarity between reference and enhanced speech using self-supervised speech representation, with the range of [0, 1].
	\item Word error rate (WER) or character error rate (CER), depending on whether the dataset is English-only or multilingual. For URGENT Challenge 2025 non-blind test set, we adopt \texttt{OWSM v3.1}{\footnote{https://huggingface.co/espnet/owsm\_v3.1\_ebf.}} following the evaluation procedure of~{\cite{zhang24h_interspeech}} to measure CER. For other benchmarks, due to the absence of reference transcripts, we use ASR transcriptions of clean speech as pseudo-labels to measure WER, which is based on \texttt{Whisper-Large-v3}{\footnote{https://github.com/openai/whisper.}}.
\end{itemize} 
\subsubsection{Non-intrusive Metrics}\label{sec:non-intrusive}
Non-intrusive metrics aim to predict human perceptual quality on a standard Mean Opinion Score (MOS) without relying on ground-truth (GT) reference. 
\begin{itemize}
	\item DNSMOS-OVRL~{\cite{reddy2022dnsmos}}: DNSMOS-OVRL is a non-intrusive speech quality predictor under the ITU-T P.835 framework, which can evaluate overall speech perceptual quality.
	\item UTMOS~{\cite{saeki2022utmos}}: UTMOS leverages fine-tuned self-supervised speech representations and ensemble to estimate human-perceived speech quality.
	\item DiMOS~{\cite{stahl2025distillation}}: DiMOS is a speech quality predictor that employs knowledge distillation on scalable SSL to achieve robust MOS estimation. 
	\item WVMOS~{\cite{andreev2023hifipp}}: WVMOS is based on fine-tuned \texttt{wav2vec 2.0} features, exhibiting strong correlations with human perceptual ratings under various acoustic distortions.
	\item SCOREQ~{\cite{ragano2024scoreq}}: SCOREQ is a speech quality evaluation model that integrates contrastive learning with regression to improve generalization across different acoustic conditions.
\end{itemize}
For all the objective metrics mentioned above, except WER and CER, higher scores indicate better performance, whereas lower WER and CER values are preferred. In addition to objective metrics, we conduct two types of subjective A/B preference tests, namely Absolute Preference Test and Similarity Preference Test. A total of 22 participants with a background in audio signal processing are recruited, each receiving 180 RMB as compensation for their participation. The former assesses the overall perceptual quality without providing clean reference speech and testing clips are randomly selected from URGENT Challenge 2025 \textit{blind} test set, whereas the latter evaluates the similarity to the clean reference in terms of linguistic content, speaker characteristics, and other acoustic details, and testing clips are selected from DNS-Challenge 2020 \textit{with reverb} test set. Each test comprises 20 listening trials, where participants are instructed to select the preferred sample from two enhanced speech candidates or choose ``Equal'' if no clear preference exists. For the Similarity Preference, the corresponding clean reference is additionally provided. To mitigate the possible selection bias, the trial order and A/B labels are randomized. Statistical significance is assessed using one-sided Kolmogorov--Smirnov (KS) tests, with $p<0.05$ considered statistically significant.
\setcounter{figure}{9}
\begin{figure*}[t]
	\centering
	\includegraphics[width=0.85\textwidth]{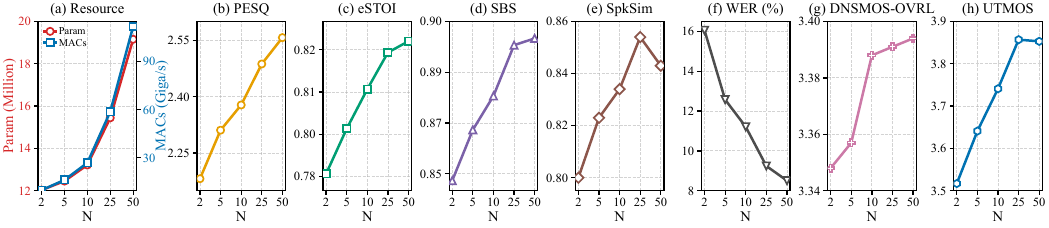}\vspace{-0.15cm}
	\caption{Metric performance under different number of Mel subbands. Results are calculated and averaged upon the DNS-Challenge 2020 \textit{without reverb} and \textit{with reverb} test sets.
	}
	\label{fig:metric-subbands}
	\vspace{-0.40cm}
\end{figure*}
\vspace{-0.27cm}
\subsection{Miscellaneous Configurations}\label{overall-configurations}
\subsubsection{Feature Configurations} For Mel extraction, all audio clips are resampled to 16 kHz. The window size is set to 1280, with 25\% shift between adjacent frames. We use a 1280 FFT and 100 Mel-frequency bins ($F_{m} = 100$), $\left\{f_{min}, f_{max}\right\}$ are set to 0 and 8 kHz, respectively. We use the mean and standard deviation of log-Mel spectra from 50 k clean speech clips and use these statistics to standardize the log-Mel inputs, and observe that it notably stabilizes the training process{\footnote{https://github.com/2ineddine/MatchaTTS-Implementation-Analysis?}}.
\subsubsection{Network Configurations} For SaT network, we use $N = 25$ subbands in Mel division, and the number of feature channels $C$ is set to 256 by default. $P=6$ DPBs are utilized. Within each DPB, the hidden feature sizes of both MHSA and FFN are both 256, and 4 attention heads are adopted. For AdaLN-SOLA, the rank is set to 8. As a result, the number of trainable parameters of SaT is 11.2 M{\footnote{Note that here we exclude the parameters of the fine-tuned WavLM in calculation of the SaT backbone.}}. For the adopted vocoder, similar to previous setup of DegVoC, totally 24 subbands are divided, and the number of channels is set to 256. Totally 12 LKCABs are adopted. As a result, the number of trainable parameters for DegVoC is 5.8 M. 
\subsubsection{Training Configurations} All experiments are based on 1 NVIDIA A100-80G GPU. For standard FM training, the batch size is 8 and the 
learning rate is initialized at 1e-4. The AdamW optimizer~{\cite{loshchilov2019decoupled}} is employed, with $\left\{\beta_{1}, \beta_{2}\right\}=\left\{0.8, 0.99\right\}$ and weight decay of 0.01. The network is updated for 1 M steps in total. For OS-FPA, the batch size is set to 4 and the learning rate is initialized at 1e-5 and we fine-tune the network for 60 k steps. For the loss weights, $\left\{\lambda_{amp}, \lambda_{mel}, \lambda_{ori}, \lambda_{perc}\right\}$ are empirically set to $\left\{40.0, 40.0, 40.0, 25.0\right\}$, respectively, to balance fidelity and perceptual quality. For both training stages, we track an exponential moving average of the DNN weights with a decay of 0.999, to be used for sampling.
\setcounter{figure}{8}
\begin{figure}[t]
	\centering
	\includegraphics[width=0.989\columnwidth]{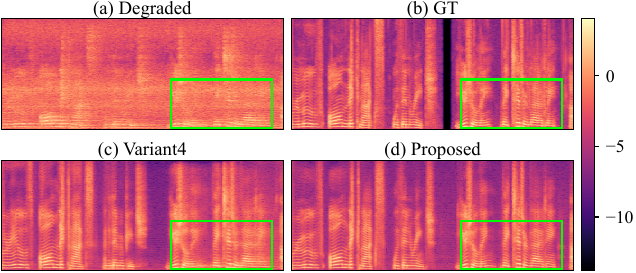}
	\vspace{-0.15cm}
	\caption{Spectral visualizations when the LKCAB and the proposed DPB are adopted. (a) Degraded speech, including noise and reverberation. (b) GT. (c) Variant4, where the LKCAB serves as the basic unit. (d) The DPB is adopted. The audio file is from DNS-Challenge 2020 \textit{with reverb} test set. 
	}
	\label{fig:spec-lkcab-dpb}
	\vspace{-0.40cm}
\end{figure} 
\setcounter{figure}{10}
\begin{figure}[t]
	\centering
	\includegraphics[width=0.96\columnwidth]{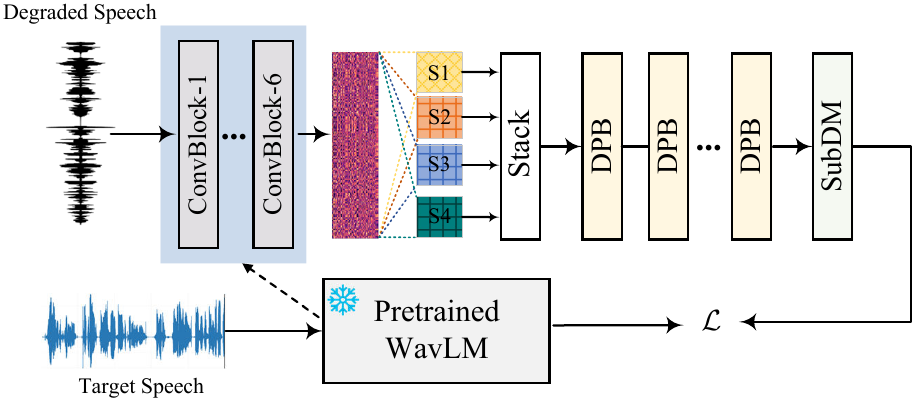}
	\vspace{-0.15cm}
	\caption{Overall diagram of the adopted SubNet, where the degraded speech first passes consecutive ConvBlocks, and the output latent feature is projected into several subspaces, followed by subspace modeling and decoding. The output is aligned with the final-layer semantic representations of the frozen pretrained WavLM given the target speech serving as the input.
	}
	\label{fig:bssl}
	\vspace{-0.58cm}
\end{figure}
\begin{table*}[t]
	\centering
	\large
	\caption{Impact of the proposed Quality-guided dual-phase vocoding strategy on DNS-Challenge 2020 \textit{without reverb} test set.}
	\label{tbl:qdv_no_reverb}
	\resizebox{0.95\textwidth}{!}{%
		\begin{tabular}{cl*{5}{c}*{5}{c}c}
			\hline\hline
			\multirow{2}{*}{\textbf{Ids}} &\multirow{2}{*}{\textbf{NIEs}} &\textbf{DNSMOS}$\uparrow$ & \multirow{2}{*}{\textbf{UTMOS}$\uparrow$} &\multirow{2}{*}{\textbf{DiMOS}$\uparrow$} & \multirow{2}{*}{\textbf{WVMOS}$\uparrow$} &\multirow{2}{*}{\textbf{SCOREQ}$\uparrow$} & \multirow{2}{*}{\textbf{PESQ}$\uparrow$} &\multirow{2}{*}{\textbf{eSTOI}$\uparrow$} & \multirow{2}{*}{\textbf{LPS}$\uparrow$} &\multirow{2}{*}{\textbf{SpkSim}$\uparrow$} &\textbf{WER}$\downarrow$ &\textbf{Replace Ratio} \\
			& &\textbf{OVRL} & & & & & & & & & \textbf{(in\%)} &\textbf{(in \%)}\\
			\hline
			1 & DNSMOS & \textbf{3.403} & 4.058 & 4.646 & 3.783 & 4.379 & 3.199 & 0.933 & 0.973 & 0.943 & 1.883 & 54.00 \\
			2 & UTMOS & \underline{3.389} & \textbf{4.124} & 4.636 & 3.816 & 4.432 & 3.295 & \underline{0.944} & 0.976 & 0.948 &1.783 & 79.33 \\
			3 & DiMOS & 3.388 & 4.046 & \textbf{4.659} & 3.757 & 4.364 & 3.137 & 0.928 & 0.972 & 0.939 & 1.959 & 36.67 \\
			4 & WVMOS & 3.388 & 4.105 & 4.643 & \textbf{3.851} & 4.444 & 3.303 & \underline{0.944} & \underline{0.977} & \underline{0.951} &\underline{1.733} & 81.33 \\
			5 & SCOREQ & 3.382 & 4.107 & 4.636 & \underline{3.834} & \textbf{4.458} & \textbf{3.445} & \textbf{0.949} & \textbf{0.978} & \textbf{0.956} & \textbf{1.657} & 98.00 \\
			\hline
			6 & Zero-phase & 3.386 & 4.004 & \underline{4.651} & 3.719 & 4.298 & 3.031 & 0.916 & 0.969 & 0.928 & 1.783 & 0.00 \\
			7 & Degraded-phase & 3.386 & \underline{4.110} & 4.636 & 3.827 & \underline{4.456} & \underline{3.345} & \textbf{0.949} & \underline{0.977} & \textbf{0.956} &1.984 & 100.00 \\
			\hline\hline
	\end{tabular}}
	\vspace{-0.25cm}
\end{table*}

\begin{table*}[t]
	\centering
	\large
	\caption{Impact of the proposed Quality-guided dual-phase vocoding strategy on DNS-Challenge 2020 \textit{with reverb} test set.}
	\label{tbl:qdv_with_reverb}
	\resizebox{0.95\textwidth}{!}{%
		\begin{tabular}{cl*{5}{c}*{5}{c}c}
			\hline\hline
			\multirow{2}{*}{\textbf{Ids}} &\multirow{2}{*}{\textbf{NIEs}} &\textbf{DNSMOS} &\multirow{2}{*}{\textbf{UTMOS}$\uparrow$} &\multirow{2}{*}{\textbf{DiMOS}$\uparrow$} & \multirow{2}{*}{\textbf{WVMOS}$\uparrow$} &\multirow{2}{*}{\textbf{SCOREQ}$\uparrow$} & \multirow{2}{*}{\textbf{PESQ}$\uparrow$} &\multirow{2}{*}{\textbf{eSTOI}$\uparrow$} & \multirow{2}{*}{\textbf{LPS}$\uparrow$} &\multirow{2}{*}{\textbf{SpkSim}$\uparrow$} &\textbf{WER}$\downarrow$ &\textbf{Replace Ratio} \\
			& &\textbf{OVRL} & & & & & & & & &\textbf{(in\%)} &\textbf{(in\%)}\\
			\hline
			1 & DNSMOS & \textbf{3.400} & 3.686 & 4.557 & 3.566 & 4.107 & \underline{2.275} & 0.759 & \underline{0.954} & \textbf{0.804} & 6.956 & 52.67 \\
			2 & UTMOS & \underline{3.393} & \textbf{3.946} & \underline{4.561} & 3.627 & \underline{4.162} & 2.229 & 0.758 & 0.934 & 0.789 & 6.906 & 2.67 \\
			3 & DiMOS & 3.384 & 3.726 & \textbf{4.587} & 3.562 & 4.113 & 2.270 & 0.758 & 0.935 & 0.794 & 6.931 & 53.33 \\
			4 & WVMOS & 3.384 & 3.892 & 4.557 & \textbf{3.643} & 4.151 & 2.245 & \underline{0.760} & \textbf{0.955} & 0.790 & 6.897 & 14.67 \\
			5 & SCOREQ & 3.390 & 3.904 & 4.557 & \underline{3.639} & \textbf{4.166} & 2.248 & \textbf{0.761} & 0.936 & 0.789 & \textbf{6.655} & 16.00 \\
			\hline
			6 & Zero-phase & 3.381 & \underline{3.932} & 4.559 & 3.636 & 4.141 & 2.228 & \underline{0.760} & 0.936 & 0.790 & \underline{6.680} & 0.00 \\
			7 & Degraded-phase & 3.379 & 3.479 & 4.500 & 3.506 & 4.049 & \textbf{2.301} & 0.758 & 0.935 & \underline{0.796} & 6.755 & 100.00 \\
			\hline\hline
	\end{tabular}}
\vspace{-0.35cm}
\end{table*}

\begin{table}[t]
	\centering
	\caption{Additional computational cost of QDV with different non-intrusive evaluators. The overhead includes one extra vocoder pass and two NIE evaluations.}
	\label{tbl:qdv_macs}
	\resizebox{\columnwidth}{!}{%
	\begin{tabular}{cccccc}
		\toprule
		\textbf{NIEs} &\textbf{DNSMOS} &\textbf{UTMOS} &\textbf{DiMOS} &\textbf{WVMOS} &\textbf{SCOREQ} \\
		\midrule
		$\Delta$MACS  &\multirow{2}{*}{45.5} &\multirow{2}{*}{14.3} &\multirow{2}{*}{17.9} &\multirow{2}{*}{13.9} &\multirow{2}{*}{19.4}\\
		(Giga/s) & & & & & \\
		\bottomrule
	\end{tabular}
}
\vspace{-0.45cm}
\end{table}
\begin{table*}[t]
	\centering
	\Huge
	\caption{Ablation studies \emph{w.r.t.} the effect of the proposed OS-FPA strategy. All results are calculated from the DNS-Challenge 2020 \textit{without reverb} test set. The QDV strategy is utilized by default.}
	\label{tbl:ablation-osfpa-without}
	\resizebox{0.90\textwidth}{!}{
		\begin{tabular}{cccccccccccccc}
			\hline\hline
			\multirow{2}{*}{\textbf{Ids}} &\textbf{Fidelity} &\textbf{Perceptual}   & 
			\multirow{2}{*}{\textbf{PESQ}$\uparrow$} &\multirow{2}{*}{\textbf{eSTOI}$\uparrow$} &\multirow{2}{*}{\textbf{LPS}$\uparrow$} &\multirow{2}{*}{\textbf{SBS}$\uparrow$} &\multirow{2}{*}{\textbf{SpkSim}$\uparrow$} & 
			\textbf{DNSMOS}$\uparrow$ &\multirow{2}{*}{\textbf{UTMOS}$\uparrow$} &\multirow{2}{*}{\textbf{DiMOS}$\uparrow$} &\multirow{2}{*}{\textbf{WVMOS}$\uparrow$} &\multirow{2}{*}{\textbf{SCOREQ}$\uparrow$} & \multirow{2}{*}{\textbf{WER (in \%)$\downarrow$}} \\
			&\textbf{Domain} &\textbf{SSL} & & & & & & \textbf{OVRL} & & & & &\\
			\hline
			1 &\xmark &\xmark  &3.321 &0.947 &0.976 &0.955 &0.939 &3.354 &4.098 &4.588 &3.803 &4.341 &2.008 \\
			2 &\xmark &WavLM-Large &3.378 &0.949 &\underline{0.978} &\underline{0.959} &0.948 &\textbf{3.390} &\textbf{4.190} &\textbf{4.664} &\textbf{3.983} &\textbf{4.577} &1.833\\
			3 &\cmark &\xmark &\underline{3.429} &0.950 &\underline{0.978} &\textbf{0.960} &\textbf{0.958} &3.378 &4.091 &4.576 &3.870 &4.453 &1.683\\
			4 &\cmark &WavLM-Large &\textbf{3.430} &\underline{0.951} &\textbf{0.979} &\textbf{0.960} &\underline{0.957} &\underline{3.387} &\underline{4.119} &\underline{4.603} &\underline{3.933} &\textbf{4.501} &\underline{1.607} \\
			5 &\cmark &Wav2vec 2.0-Large &3.422 &\textbf{0.952} &\underline{0.978} &\textbf{0.960} &\underline{0.957} &\underline{3.387} &4.117 &4.590 &3.927 & 4.492 &\textbf{1.482} \\
			6 &\cmark &HuBERT-Large &3.426 &\textbf{0.952} &\underline{0.978} &\textbf{0.960} &\underline{0.957} &3.386 &4.108 &4.585 &3.913 &4.485 &1.683\\
			\hline\hline
		\end{tabular}
	}
	\vspace{-0.15cm}
\end{table*}
\begin{table*}[t]
	\centering
	\Huge
	\caption{Ablation studies \emph{w.r.t.} the effect of the proposed OS-FPA strategy. All results are calculated from the DNS-Challenge 2020 \textit{with reverb} test set. The QDV strategy is utilized by default.}
	\label{tbl:ablation-osfpa-with}
	\resizebox{0.90\textwidth}{!}{
		\begin{tabular}{cccccccccccccc}
			\hline\hline
			\multirow{2}{*}{\textbf{Ids}} &\textbf{Fidelity} &\textbf{Perceptual}   & 
			\multirow{2}{*}{\textbf{PESQ}$\uparrow$} &\multirow{2}{*}{\textbf{eSTOI}$\uparrow$} &\multirow{2}{*}{\textbf{LPS}$\uparrow$} &\multirow{2}{*}{\textbf{SBS}$\uparrow$} &\multirow{2}{*}{\textbf{SpkSim}$\uparrow$} & 
			\textbf{DNSMOS}$\uparrow$ &\multirow{2}{*}{\textbf{UTMOS}}$\uparrow$ &\multirow{2}{*}{\textbf{DiMOS}$\uparrow$} &\multirow{2}{*}{\textbf{WVMOS}$\uparrow$} &\multirow{2}{*}{\textbf{SCOREQ}$\uparrow$} & \multirow{2}{*}{\textbf{WER (in \%)$\downarrow$}} \\
			&\textbf{Domain} &\textbf{SSL} & & & & & & \textbf{OVRL} & & & &  &\\
			\hline
			1 &\xmark &\xmark  &2.293 &0.752 &0.939 &0.903 &0.790 &3.300 &3.889 &4.409 &3.539 &3.925 &7.050 \\
			2 &\xmark &WavLM-Large &2.330 &0.768 &\textbf{0.942} &0.911 &0.807 &\textbf{3.399} &\textbf{4.159} &\textbf{4.625} &\textbf{3.914} &\textbf{4.515} &6.931\\
			3 &\cmark &\xmark &2.340 &\underline{0.770} &\textbf{0.942} &0.911 &0.839 &3.381 &3.950 &4.469 &3.732 &4.235 &\underline{6.605} \\
			4 &\cmark &WavLM-Large &\textbf{2.357} &0.769 &\textbf{0.942} &\textbf{0.913} &\textbf{0.854} &\underline{3.394} &4.039 &\underline{4.536} &3.874 &4.360 &\textbf{6.429} \\
			5 &\cmark &Wav2vec 2.0-Large &\underline{2.356} &\underline{0.770} &\underline{0.941} &\underline{0.912} &0.841 &\textbf{3.399} &\underline{4.052} &4.518 &\underline{3.893} &\underline{4.360} &6.755 \\
			6 &\cmark &HuBERT-Large &2.351 &\textbf{0.771} &\underline{0.941} &\textbf{0.913} &\underline{0.852} &\underline{3.394} &4.025 &4.499 &3.852 &4.340 &6.655\\
			\hline\hline
		\end{tabular}
	}
	\vspace{-0.15cm}
\end{table*}
\renewcommand\arraystretch{0.9}
\begin{table*}[t]
	\centering
	\huge
	\caption{Objective comparisons with multiple advanced SE models on the DNS-Challenge 2020 \textit{without reverb} test set. ``Type'' denotes the type of SE methods. ``-'' denotes no score is reported due to the lack of pretrained model weights.}
	\label{tbl:objective_comparisons-dns-wo-reverb}
	\resizebox{0.90\textwidth}{!}{
		\begin{tabular}{cccccccccccccc}
			\hline\hline
			\multirow{3}{*}{\textbf{Ids}} &
			\multirow{3}{*}{\textbf{Models}} &
			\multirow{3}{*}{\textbf{Type}} &
			\multicolumn{6}{c}{\textbf{Intrusive Metrics}} &
			\multicolumn{5}{c}{\textbf{Non-intrusive Metrics}} \\
			\cmidrule(lr){4-9}\cmidrule(lr){10-14}
			& & &
			\multirow{2}{*}{\textbf{PESQ$\uparrow$}} &
			\multirow{2}{*}{\textbf{eSTOI$\uparrow$}} &
			\multirow{2}{*}{\textbf{LPS$\uparrow$}} &
			\multirow{2}{*}{\textbf{SBS$\uparrow$}} &
			\multirow{2}{*}{\textbf{SpkSim$\uparrow$}} 
			&\multirow{2}{*}{\textbf{WER (in \%)$\downarrow$}} &
			\textbf{DNSMOS$\uparrow$} &
			\multirow{2}{*}{\textbf{UTMOS$\uparrow$}} &
			\multirow{2}{*}{\textbf{DiMOS$\uparrow$}} &
			\multirow{2}{*}{\textbf{WVMOS$\uparrow$}} &
			\multirow{2}{*}{\textbf{SCOREQ$\uparrow$}} 
			 \\
			& & & & & & & & &\textbf{OVRL}
			 & & & & \\
			\hline
			1 &Degraded &- &1.582 &0.810 &0.898 &0.882 &0.945 &3.616 &2.483 &2.365 &3.064 &1.788 &2.789 \\
			\hline
			2 &TFGridNet &D    & 3.180 & 0.943 & 0.966 & 0.936 & 0.938 & 2.938 & 3.310 & 3.832 & 4.350 & 3.571 & 4.150  \\
			3 & BSRNN &D          &\underline{3.347} &\textbf{0.954} & 0.971 & 0.946 &\underline{0.945}  & 3.064 & 3.335 & 3.984 & 4.517 & 3.733 & 4.373  \\
			4 &StoRM  &G         & 2.695 & 0.923 & 0.955 & 0.926 & 0.935 & 4.074 & 3.290 & 3.725 & 4.389 & 3.523 & 3.992  \\
			5 &VoiceFixer &G     & 1.923 & 0.742 & 0.900 & 0.846 & 0.469 & 7.389 & 3.249 & 3.107 & 4.056 & 2.887 & 3.550  \\
			6 &CleanMel &G       & 3.077 & 0.932 & 0.961 & 0.941 & 0.933 & 2.809  & 3.401 & 3.686 & 4.564 & 3.798 & 4.400 \\
			7 &FlowSE  &G       & 2.168 & 0.859 & 0.911 & 0.888 & 0.828 & 8.578 & 3.274 & 3.072 & 4.196 & 3.010 & 3.895  \\
			8 &SenSE  &G        & 2.809 & 0.925 & 0.966 & 0.939 & 0.931 & 4.319 & 3.375 & 3.850 & \underline{4.659} & 3.865 &\textbf{4.592}  \\
			9 &StuPASE  &G      & 2.508 & 0.883 & 0.967 & 0.936 & 0.897 &\underline{2.075} & 3.393 & 3.912 & 4.657 & 3.740 & 4.483  \\
			10 &SELM     &G      & - & - & - & - & - & - & 3.258 & - & - & - & -  \\
			11 &MaskSR   &G      & - & - & - & - & - & - & 3.339 & - & - & - & -  \\
			12 &MaskSR2  &G      & - & - & - & - & - & - & 3.400 & - & - & - & -  \\
			13 &LLaSE-G1  &G     & 1.342 & 0.658 & 0.896 & 0.849 & 0.602 & 8.451 & 3.138 & 3.237 & 4.379 & 3.164 & 3.967  \\
			14 & Metis  &G        & 1.279 & 0.748 & 0.940 & 0.894 & 0.493 & 5.424 &\textbf{3.434} & \textbf{4.195} &\textbf{4.684} & 4.029 & 4.617  \\
			15 &AnyEnhance  &G   & 2.949 & 0.909 & 0.961 & 0.929 & 0.682 & 4.420 & 3.416 & 3.955 & 4.487 & 3.755 & 4.481  \\
			16 &GenSE   &G       & 1.254 & 0.593 & 0.819 & 0.834 & 0.769 & 14.867 & 3.422 & 3.924 & 4.663 & 3.710 & 4.285  \\
			17 &UniSE   &G       & 1.969 & 0.833 & 0.926 & 0.904 & 0.771 & 7.559 & 3.424 & 4.057 & 4.588 &\textbf{4.049} & 4.464  \\
			18 &L3-SE    &G      & 2.534 & 0.875 & 0.959 & 0.930 & 0.837 & 4.219 &\underline{3.432} & 3.859 & 4.639 & 3.854 & 4.474  \\
			19 &Xingchen-UniFE &G  & 2.396 & 0.859 & 0.952 & 0.929 & 0.838 & 4.721 & 3.423 & 4.115 & 4.653 &\underline{4.042} & 4.526  \\
			20 & PASE    &G       & 2.665 & 0.903 & 0.972 & 0.944 & 0.633  & 2.129 & 3.381 & 4.009 & 4.495 & 3.967 & 4.491 \\
			21 & PhASE-Flow  &G   & 2.973 & 0.925 & 0.970 & 0.945 & 0.854 & 2.708 & 3.398 & 4.110 & 4.581 & 3.908 & 4.577  \\
			22 &UniPASE  &G      & 3.019 & 0.929 &\underline{0.974} &\underline{0.950} & 0.691 & 2.509 & 3.368 & 4.041 & 4.518 & 3.999 &\underline{4.588}  \\
			\hline
			23 &\textbf{BaME (Pro.)} &G
			& \textbf{3.430}
			& \underline{0.951}
			& \textbf{0.979}
			& \textbf{0.960}
			& \textbf{0.957}
			& \textbf{1.607}
			& 3.387
			&\underline{4.119}
			& 4.603
			& 3.933
			& 4.501
			 \\
			\hline\hline
		\end{tabular}
	}
	\vspace{-0.40cm}
\end{table*}
\begin{figure*}[t]
	\centering
	\includegraphics[width=0.95\textwidth]{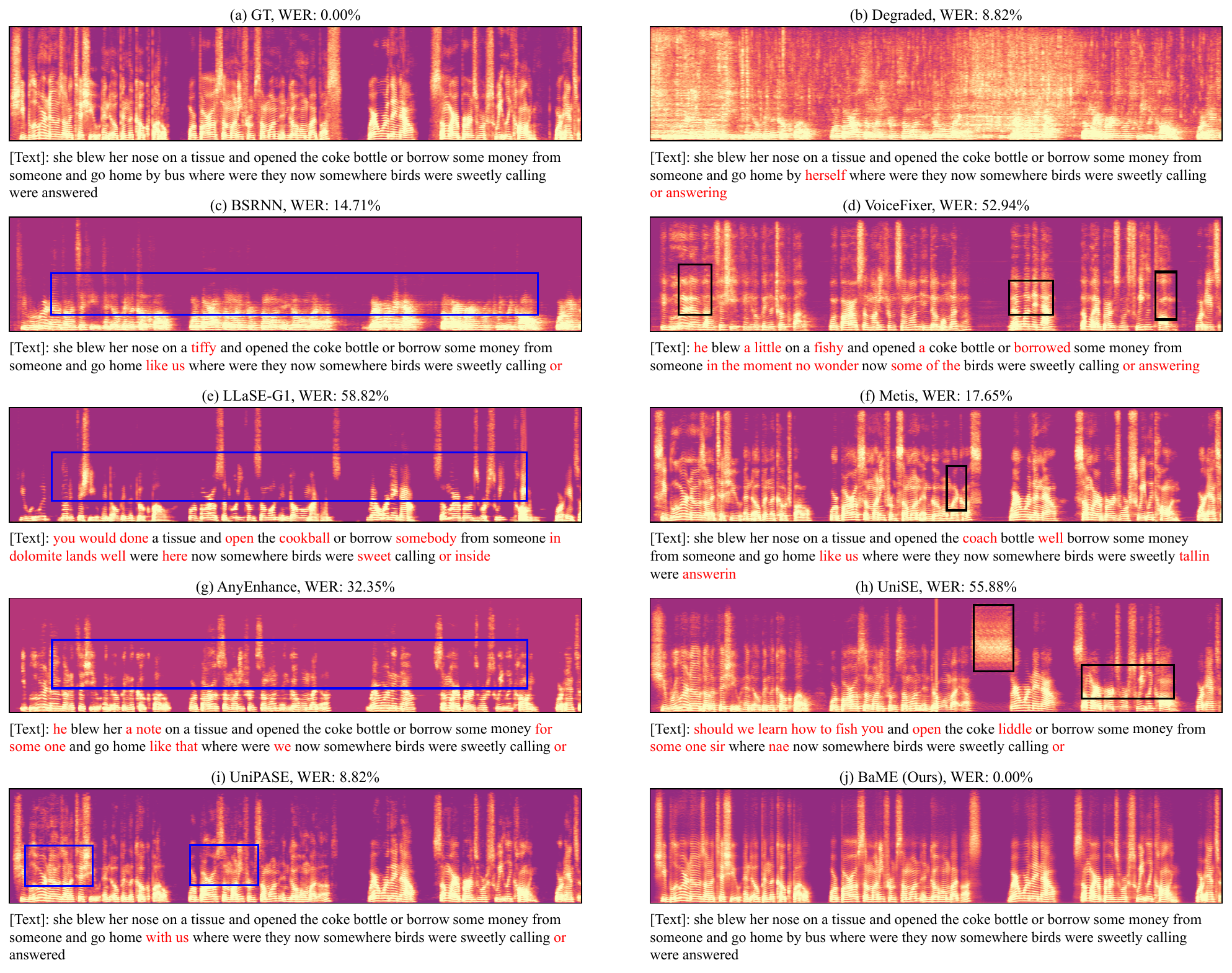}
	\vspace{-0.15cm}
	\caption{Spectral visualizations among different SE methods. The audio file is selected from DNS-Challenge 2020 \textit{with reverb} test set. Words highlighted in \textcolor{red}{red} indicate transcription errors. 
	}
	\label{fig:sota_spectra_comparisons}
	\vspace{-0.45cm}
\end{figure*} 
\renewcommand\arraystretch{0.9}
\begin{table*}[t]
	\centering
	\huge
	\caption{Objective comparisons with multiple advanced SE models on the DNS-Challenge 2020 \textit{with reverb} test set. ``-'' denotes no score is reported due to the lack of pretrained model weights.}
	\label{tbl:objective_comparisons-dns-wi-reverb}
	\resizebox{0.92\textwidth}{!}{
		\begin{tabular}{cccccccccccccc}
			\hline\hline
			\multirow{3}{*}{\textbf{Ids}} &
			\multirow{3}{*}{\textbf{Models}} &
			\multirow{3}{*}{\textbf{Type}} &
			\multicolumn{6}{c}{\textbf{Intrusive Metrics}} &
			\multicolumn{5}{c}{\textbf{Non-intrusive Metrics}} \\
			\cmidrule(lr){4-9}\cmidrule(lr){10-14}
			& & &
			\multirow{2}{*}{\textbf{PESQ$\uparrow$}} &
			\multirow{2}{*}{\textbf{eSTOI$\uparrow$}} &
			\multirow{2}{*}{\textbf{LPS$\uparrow$}} &
			\multirow{2}{*}{\textbf{SBS$\uparrow$}} &
			\multirow{2}{*}{\textbf{SpkSim$\uparrow$}} &
			\multirow{2}{*}{\textbf{WER (in \%)$\downarrow$}} &
			\textbf{DNSMOS$\uparrow$} &
			\multirow{2}{*}{\textbf{UTMOS$\uparrow$}} &
			\multirow{2}{*}{\textbf{DiMOS$\uparrow$}} &
			\multirow{2}{*}{\textbf{WVMOS$\uparrow$}} &
			\multirow{2}{*}{\textbf{SCOREQ$\uparrow$}}  \\
			& & & & & & & & 
			&\textbf{OVRL} & & & &  \\
			\hline
			1 &Degraded &- &1.164 &0.348 &0.633 &0.739 &0.699 &9.669 &1.392 &1.300 &2.321 &-0.679 &2.105 \\
			\hline
			2 & TFGridNet &D   & 1.509 & 0.494 & 0.878 & 0.837 & 0.695 & 8.764 & 2.603 & 1.420 & 3.001 & 2.343 & 2.785  \\
			3 &BSRNN  &D       & 1.832 & 0.640 & 0.819 & 0.816 & 0.678 & 18.584 & 2.913 & 2.294 & 3.158 & 2.543 & 2.683  \\
			4 &StoRM  &G       & 1.367 & 0.493 & 0.594 & 0.699 & 0.585 & 48.001 & 2.866 & 1.829 & 3.140 & 1.874 & 2.139  \\
			5 &VoiceFixer &G    & 1.387 & 0.638 & 0.671 & 0.726 & 0.315 & 30.305 & 3.136 & 2.310 & 3.569 & 2.420 & 3.550  \\
			6 &CleanMel   &G    &\textbf{2.599} &\textbf{0.771} &\underline{0.936} & \underline{0.897} &\underline{0.839} &\textbf{6.210}  &3.285 & 3.378 & 4.427 & 3.433 & 2.744 \\
			7 &FlowSE   &G      & 1.269 & 0.370 & 0.708 & 0.742 & 0.641 & 26.670 & 2.285 & 1.350 & 2.818 & 1.194 & 2.645  \\
			8 &SenSE    &G      & 1.836 & 0.694 & 0.918 & 0.884 & 0.780 & 10.171 & 3.367 & 3.556 &4.588 & 3.617 & 4.410  \\
			9 &StuPASE   &G     & 1.764 & 0.765 & 0.921 & 0.889 & 0.725 & 8.513 & 3.360 & 3.911 & 4.630 & 3.696 & 4.445  \\
			10 &SELM    &G       & - & - & - & - & - & - & 2.695 & - & - & - & -  \\
			11 &MaskSR   &G      & - & - & - & - & - & -  & 3.253 & - & - & - & - \\
			12 &LLaSE-G1   &G    & 1.096 & 0.520 & 0.676 & 0.748 & 0.385  & 38.841 & 3.049 & 2.347 & 3.896 & 2.647 & 3.072 \\
			13 &Metis    &G      & 1.207 & 0.594 & 0.857 & 0.848 & 0.456 & 20.568 &\textbf{3.430} &\textbf{4.152} &\textbf{4.625} &\textbf{3.995} &\textbf{4.528}  \\
			14 &AnyEnhance   &G  & 1.794 & 0.647 & 0.871 & 0.846 & 0.719  & 14.139 & 3.201 & 2.747 & 4.102 & 2.840 & 3.584 \\
			15 &GenSE    &G      & 1.126 & 0.305 & 0.513 & 0.680 & 0.516 & 59.543 & 3.173 & 2.085 & 3.708 & 2.672 & 2.684  \\
			16 &UniSE    &G      & 1.464 & 0.629 & 0.742 & 0.804 & 0.593 & 36.188 & 3.415 & 3.780 & 4.398 & 3.715 & 4.174  \\
			17 &L3-SE     &G     & 1.671 & 0.735 & 0.916 & 0.880 & 0.662 & 8.287 &\underline{3.426} & 3.594 &\underline{4.604} & 3.817 & 4.163  \\
			18 &Xingchen-UniFE   &G & 1.921 & 0.700 & 0.874 & 0.869 & 0.716  & 16.349 & 3.289 & 3.790 & 4.499 & 3.727 & 4.161 \\
			19 &VoiceBridge  &G  & 1.583 & - & - & - & - &- & 3.318 & 3.382 & - & - & -   \\
			20 &PASE    &G       & 1.418 & 0.475 & 0.905 & 0.860 & 0.589 & 10.095 & 2.701 & 1.636 & 3.303 & 2.953 & 3.289  \\
			21 &PhASE-Flow   &G  & 1.765 & 0.764 & 0.895 & 0.875 & 0.593 & 13.111 & 3.365 & 3.807 & 4.455 & 3.888 & 4.298  \\
			22 &UniPASE   &G     & 1.742 & 0.761 & 0.931 & 0.896 & 0.582 & 8.162 & 3.310 & 3.619 & 4.418 &\underline{3.946} &\underline{4.392}  \\
		    \hline
			23 &\textbf{BaME (Pro.)}  &G
			&\underline{2.357} &\underline{0.769} & \textbf{0.942} & \textbf{0.913}
			& \textbf{0.854} &\underline{6.429} & 3.394 &\underline{4.054} & 4.536
			& 3.874 & 4.360  \\
			\hline\hline
		\end{tabular}
	}
	\vspace{-0.40cm}
\end{table*}
\renewcommand\arraystretch{0.9}
\begin{table*}[t]
	\centering
	\huge
	\caption{Objective comparisons with advanced SE methods on the EARS-WHAM\_v2 denoising and EARS-Reverb\_v2 dereverberation benchmarks. ``D'' and ``G'' denote discriminative and generative methods, respectively. ``-'' indicates unavailable results.}
	\label{tbl:objective_comparisons_ears}
	\resizebox{0.92\textwidth}{!}{
		\begin{tabular}{c l c ccccc ccccc}
			\hline\hline
			\multirow{2}{*}{\textbf{Ids}} &
			\multirow{2}{*}{\textbf{Models}} &
			\multirow{2}{*}{\textbf{Type}} &
			\multicolumn{5}{c}{\textbf{EARS-WHAM\_v2 (Denoising)}} &
			\multicolumn{5}{c}{\textbf{EARS-Reverb\_v2 (Dereverberation)}} \\
			\cmidrule(lr){4-8}\cmidrule(lr){9-13}
			& & &
			\textbf{PESQ$\uparrow$} & \textbf{eSTOI$\uparrow$} &
			\textbf{WER(\%)$\downarrow$} & \textbf{DiMOS$\uparrow$} &
			\textbf{SCOREQ$\uparrow$} &
			\textbf{PESQ$\uparrow$} & \textbf{eSTOI$\uparrow$} &
			\textbf{WER (in \%)$\downarrow$} & \textbf{DiMOS$\uparrow$} &
			\textbf{SCOREQ$\uparrow$} \\
			\hline
			1 &Degraded & - & 1.243 & 0.640 & 4.666 & 2.468 & 2.108 & 1.321 & 0.584 & 1.950 & 3.021 & 3.116 \\
			\hline
			2 & TFGridNet & D & 2.343 & 0.815 & 10.643 & 3.834 & 3.607 & 1.654 & 0.690 & 1.937 & 3.370 & 3.363 \\
			3 & BSRNN & D & 2.360 & 0.822 & 8.529 & 3.768 & 3.429 & 1.993 & 0.772 & 4.713 & 3.616 & 3.278 \\
			4 & StoRM & G & 1.926 & 0.774 & 14.350 & 4.030 & 3.328 & 2.371 & 0.846 & 1.995 & 4.055 & 3.379 \\
			5 & VoiceFixer & G & 1.588 & 0.642 & 16.000 & 3.411 & 2.827 & 1.561 & 0.655 & 5.940 & 3.495 & 2.771 \\
			6 & CleanMel & G & \underline{2.531} & 0.818 & 6.357 & 4.297 & 3.908 & \underline{2.697} & \underline{0.851} & \textbf{1.122} & 4.281 & 3.591 \\
			7 & FlowSE & G & 1.884 & 0.754 & 16.659 & 4.063 & 3.518 & 1.335 & 0.560 & 4.034 & 2.981 & 2.941 \\
			8 &SenSE &G &2.150 &0.768 &12.535 &4.190 &4.050 &2.251 &0.798 &2.473 &4.388 &3.907 \\
			9 & StuPASE & G & 2.073 & 0.777 & 7.057 & 4.331 & 4.295 & 1.954 & 0.760 & 3.024 & 4.407 & 4.123 \\
			10 & LLaSE-G1 & G & 1.291 & 0.581 & 17.334 & 3.893 & 3.353 & 1.172 & 0.492 & 11.363 & 3.809 & 3.248 \\
			11 & Metis & G & 1.286 & 0.669 & 12.987 & \textbf{4.559} & \textbf{4.523} & 1.275 & 0.672 & 4.413 & \textbf{4.530} & \textbf{4.488} \\
			12 & GenSE & G & 1.360 & 0.513 & 47.035 & 4.021 & 3.006 & 1.217 & 0.430 & 44.277 & 3.549 & 2.361 \\
			13 & UniSE & G & 1.659 & 0.816 & 21.793 & \underline{4.401} & 4.240 & 1.603 & 0.732 & 7.912 & 4.240 & 4.222 \\
			14 & PASE & G & 2.200 & 0.816 & \textbf{4.761} & 4.260 & 3.980 & 1.514 & 0.635 & 3.167 & 3.391 & 3.276 \\
			15 & UniPASE & G & 2.406 & \underline{0.846} & 3.153 & 4.249 & 4.164 & 1.878 & 0.730 & 1.568 & 4.221 & 4.204 \\
			\hline
			16 & \textbf{BaME (Pro.)} & G &
			\textbf{2.673} & \textbf{0.847} & 6.272 &
			4.354 & \underline{4.325} &
			\textbf{2.807} & \textbf{0.869} & \underline{1.227} &
			\underline{4.502} & \underline{4.322} \\
			\hline\hline
		\end{tabular}
	}
	\vspace{-0.40cm}
\end{table*}
\begin{table*}[t]
	\centering
	\huge
	\caption{Objective comparisons with multiple advanced SE models on the GSR test set.}
	\label{tbl:objective_comparisons-gsr}
	\resizebox{0.92\textwidth}{!}{
		\begin{tabular}{cccccccccccccc}
			\hline\hline
			\multirow{3}{*}{\textbf{Ids}} &
			\multirow{3}{*}{\textbf{Models}} &
			\multirow{3}{*}{\textbf{Type}} &
			\multicolumn{6}{c}{\textbf{Intrusive Metrics}} &
			\multicolumn{5}{c}{\textbf{Non-intrusive Metrics}} \\
			\cmidrule(lr){4-9}\cmidrule(lr){10-14}
			& & &
			\multirow{2}{*}{\textbf{PESQ$\uparrow$}} &
			\multirow{2}{*}{\textbf{eSTOI$\uparrow$}} &
			\multirow{2}{*}{\textbf{LPS$\uparrow$}} &
			\multirow{2}{*}{\textbf{SBS$\uparrow$}} &
			\multirow{2}{*}{\textbf{SpkSim$\uparrow$}} &
			\multirow{2}{*}{\textbf{WER (in \%)$\downarrow$}} &
			\textbf{DNSMOS$\uparrow$} &
			\multirow{2}{*}{\textbf{UTMOS$\uparrow$}} &
			\multirow{2}{*}{\textbf{DiMOS$\uparrow$}} &
			\multirow{2}{*}{\textbf{WVMOS$\uparrow$}} &
			\multirow{2}{*}{\textbf{SCOREQ$\uparrow$}}  \\
			& & & & & & & & 
			&\textbf{OVRL} & & & &  \\
			\hline
			1 &Degraded &- &1.943 &0.582 &0.845 &0.814 &0.780 &7.831 &2.392 &2.634 &3.206 &2.000 &3.010  \\
			\hline
			2 & TFGridNet &D  &2.457 &0.681 &0.882 &0.858 &0.781 &7.801 &2.993 &3.357 &3.157 &3.782 &3.775   \\
			3 &BSRNN  &D  &\underline{2.612} &\underline{0.698} &0.888 &0.882 &\underline{0.803} &8.156 &2.984 &3.551 &3.465 &3.868 &3.727   \\
			4 &VoiceFixer  &G     & 2.032 &0.557 &0.798 &0.835 &0.423 &12.648 &2.997 &3.432 &3.691 &3.806 &3.693   \\
			5 &Metis &G    &1.335 &0.531 &0.858 &0.826 &0.344 &16.401 &\textbf{3.287} &\textbf{4.074} &\underline{3.723} &\textbf{4.355} &\textbf{4.498}  \\
			6 &UniSE   &G    &1.759 &0.582 &0.806 &0.830 &0.531 &16.401 &3.135 &3.933 &3.622 &3.941 &4.174  \\
			7 &UniPASE   &G      & 2.488 &0.687 &\underline{0.909} &\underline{0.898} &0.582 &\underline{6.915} &3.075 &3.875 &\textbf{3.732} &4.169 &4.172  \\
			8 &\textbf{BaME (Pro.)}    &G      &\textbf{2.888} &\textbf{0.733} &\textbf{0.922} &\textbf{0.924} &\textbf{0.836} &\textbf{6.087} &\underline{3.163} &\underline{3.943} &3.661 &\underline{4.274} &\underline{4.309}  \\
			\hline\hline
		\end{tabular}
	}
	\vspace{-0.40cm}
\end{table*}
\begin{table*}[t]
	\centering
	\caption{Objective comparisons with multiple advanced SE models on the URGENT Challenge 2025 non-blind test set. ``D+G'' denotes discriminative and generative SE methods are combined. The results of degraded speech as well as that of Rank-1/Rank-2/Rank-3 are directly used from the challenge leaderboard.}
	\label{tbl:objective_comparisons-urgent}
	\resizebox{0.90\textwidth}{!}{
		\begin{tabular}{ccccccccccc}
			\hline\hline
			\textbf{Ids} &\textbf{Models} &\textbf{Type} &\textbf{PESQ$\uparrow$} &\textbf{eSTOI$\uparrow$} &\textbf{LPS$\uparrow$} &\textbf{SBS$\uparrow$} &\textbf{SpkSim$\uparrow$} &\textbf{CER (in \%)$\downarrow$} &
			\textbf{DNSMOS-OVRL$\uparrow$} &\textbf{UTMOS$\uparrow$}  \\
			\hline
			1 &Degraded &- &1.37 &0.61 &0.62 &0.75 &0.63 &18.71 &1.84 &1.56  \\
			\hline
			2 & Rank-1 &D  &\textbf{2.95} &\textbf{0.86} &\textbf{0.86} &\textbf{0.91} &\textbf{0.85} &\textbf{11.08} &3.01 &2.40   \\
			3 &Rank-2  &D+G  &\underline{2.74} &\underline{0.84} &0.84 &0.89 &\underline{0.83} &\underline{12.06} &3.00 &2.31   \\
			4 &Rank-3  &D+G   &2.63 &0.82 &0.82 &0.88 &0.82 &13.85 &3.02 &2.34   \\
			\hline
			5 &TFGridNet &D &2.434 &0.801 &0.792 &0.851 &0.592 &15.044 &2.942 &2.108 \\
			6 &BSRNN & D &1.840 &0.636 &0.674 &0.780 &0.531 &26.803 &2.738 &2.226 \\
			7 &VoiceFixer &G &1.548 &0.571 &0.599 &0.746 &0.390 &34.082 &2.898 &2.057 \\
			8 &Metis &G &1.211 &0.545 &0.718 &0.800 &0.442 &22.223 &\underline{3.298} &\underline{3.112} \\
			9 &UniSE &G &1.551 &0.613 &0.601 &0.783 &0.562 &36.077 &3.281 &3.040 \\
			10 &UniPASE &G &2.113 &0.694 &\underline{0.844} &0.892 &0.799 &12.709 &3.212 &2.956 \\
			11 &\textbf{BaME (Pro.)} & G &2.259 &0.739 &0.843 &\underline{0.896} &0.778 &12.584 &\textbf{3.336} &\textbf{3.293}\\
			\hline\hline
		\end{tabular}
	}
	\vspace{-0.40cm}
\end{table*}

\vspace{-0.25cm}
\section{Experimental Results and Analysis}\label{sec:experiment-results-analysis}
\subsection{Ablation Studies}\label{sec:ablation-study}
Let us begin with a DiT trained with velocity loss, and the semantic input $\mathbf{\tilde{D}}$ is removed, whose structure configuration is similar to~{\cite{wang25s_interspeech}}. $\text{NFE} = 8$ is adopted and the proposed QDV tactic is not employed if not otherwise specified. Only noise and reverberation degradations are considered in the training process. The DNS-Challenge 2020 \textit{without reverb} and \textit{with reverb} test sets are utilized for ablation evaluations.
\subsubsection{Impact of Data Prediction Loss} From Id1 to Id2 in Table~{\ref{tbl:ablation-arch}}, replacing velocity prediction with data prediction can improve most metrics, especially those related to speech fidelity and a marginal decrease is observed in UTMOS. It indicates that it is beneficial for generative SE to improve speech fidelity by directly estimating the target Mel spectrum and data-prediction parameterization in the inference stage.
\subsubsection{Impact of Backbone Architecture}\label{sec:effect-backbone} We first compare the proposed SaT with the conventional DiT backbone. As given by Id2 and Id3 in Table~{\ref{tbl:ablation-arch}}, SaT achieves overall better performance with only 2.9\% (9.67 M vs. 331 M) of the parameters and 66.6\% of the computational complexity (6.65 vs 9.98 GMACs/s for a single sampling pass), demonstrating the efficiency of Mel subband-aware modeling. Further, we investigate four architectural variants, as shown in Fig.~{\ref{fig:backbone-variants}}. Concretely, Variant1 removes the filtering branch of CSDM, \emph{i.e.}, we directly map the enhanced Mel spectrum. Variant2 and Variant3 adopt BiGRU and BiLSTM for both frame- and subband modeling, respectively, and the FFN is removed. Variant4 replaces the DPB with the convolution-based module called LKCAB proposed in~{\cite{li2025bridgevoc}}. The channel size $C$ remains 256 for fairness. Notably, comparing Id3 and Id4, the proposed SaT consistently achieves the best metric performance, revealing the effectiveness of collaborative filtering and refinement. Besides, Id3 consistently outperforms Id5-Id6, indicating the superiority of self-attention for frame modeling, and GRU is adequate for subband modeling. Interestingly, although LKCAB has exhibited promising performance in the information generation-related task, \emph{i.e.}, neural vocoder, it is notably inferior to the adopted DPB, as shown in Id3 and Id7, as well as the spectral visualization in Fig.~{\ref{fig:spec-lkcab-dpb}}, indicating that convolution-based modeling, with its spatially shared kernels, may be more challenging in capturing feature dependencies for suppression-oriented speech restoration tasks, \emph{e.g.}, speech denoising or dereverberation.  
\subsubsection{Impact of Trajectory Modeling}\label{sec:effect-trajectory} We investigate different trajectory modeling strategies, including FM and Schrodinger bridge with variance-preserving (Sch-VP) and variance-exploding (Sch-VE) schedulers~{\cite{jukic24_interspeech}}. As given by Id3, Id8-Id9 in Table~{\ref{tbl:ablation-arch}}, FM remarkably outperforms both bridge variants across various metrics. This suggests that simple linear interpolation path of FM may be sufficient for Mel-domain generative modeling. Moreover, unlike the bridge variants that adopt degraded Mel spectra as the source distribution, FM herein starts from standard Gaussian noise. Therefore, by incorporating the structural prior of Mel-spectrum for backbone design, the choice of source distribution may be less critical than expected.
\subsubsection{Impact of Number of Subbands}\label{sec:effect-subband} We analyze the effect of the number of Mel subbands $N$, ranging from $N=2$ to $N=50$. As shown in Fig.~{\ref{fig:metric-subbands}}, as $N$ increases, most objective metrics gradually improve, indicating the benefit of fine-grained subband modeling for Mel representations. Note that the performance gains gradually diminish, especially when $N$ increases from 25 to 50, while the parameters and complexity are substantially increased. Therefore, we finally choose $N=25$ by default to achieve a favorable trade-off between performance and computational efficiency.
\subsubsection{Effect of Auxiliary Semantic Features}\label{sec:effect-semantic} We explore the impact of auxiliary semantic features on enhancement performance, as shown in Table~{\ref{tbl:semantic_ablation}}. Going from Id1 to Id2, \emph{i.e.}, incorporating denoised semantic representations extracted by the fine-tuned WavLM (DeWavLM), yields remarkable improvements across multiple metrics, particularly reducing WER by 53.81\%. This fully demonstrates the effectiveness of semantic conditioning in preserving linguistic content and mitigating speech hallucinations. 

Nonetheless, the additional WavLM can introduce substantial parameters and computational overhead. To alleviate this issue, we further explore a lightweight semantic extractor, termed SubNet, as shown in Fig.~{\ref{fig:bssl}}. To be specific, it adopts the first six convolutional blocks of WavLM initialized with pretrained weights, followed by four independent low-rank projections similar to Eq.~{(\ref{eqn:12})}. Inspired by the subband modeling paradigm of SaT, we regard each subspace as a special type of subband, processed by $P=6$ DPBs, followed by a subspace decoding module (SubDM) to estimate semantic representations. The training procedure of SubNet is similar to DeWavLM. 

Compared with Id2, SubNet reduces the parameters by 91.83\% and computational complexity by 13.50\%, while consistently outperforming the case without semantic features (Id1). Nevertheless, a noticeable performance gap still exists compared with DeWavLM. Therefore, it remains challenging to effectively extract semantic information under limited computational resources, and we still adopt DeWavLM as the default semantic extractor.
\subsubsection{Effect of the QDV Strategy}\label{sec:effect-phase} We explore the effectiveness of the proposed quality-guided dual-phase vocoding (QDV) strategy with five NIEs: DNSMOS-OVRL, UTMOS, DiMOS, WVMOS and SCOREQ. The results on DNS-Challenge 2020 \textit{without reverb} and $\textit{with reverb}$ test sets are presented in Table~{\ref{tbl:qdv_no_reverb}} and Table~{\ref{tbl:qdv_with_reverb}}, respectively. Note that the replace ratio denotes the proportion of samples for which degraded-phase reconstruction is selected.

Several observations can be made. First, simply incorporating degraded phase in the vocoder generally improves speech quality and fidelity under noise-only cases except WER. However, when reverberation is involved, degraded phase can degrade most non-intrusive scores especially UTMOS, while obtaining comparable performance in intrusive metrics. This fully illustrates the necessity of adaptive phase selection. Second, as expected, each NIE achieves the highest score on its corresponding metric due to the selection criterion. Besides, we notice that different NIEs may exhibit distinct preferences for degraded phase. To be specific, UTMOS and SCOREQ seem less likely to select degraded-phase-oriented speech generation under reverberation, whereas DNSMOS and DiMOS show less sensitivity to this change. Despite these differences, the overall performance across evaluation metrics remains relatively consistent, indicating the robustness of the QDV strategy.

As QDV requires two NIE evaluations and an additional vocoding pass, we report the resulting increased computational complexity, as shown in Table~{\ref{tbl:qdv_macs}}. Considering its favorable overall performance and moderate additional complexity, SCOREQ is adopted as the NIE by default.
\subsubsection{Effect of the OS-FPA Strategy} We finally investigate the effect of the proposed OS-FPA strategy, as shown in Table~{\ref{tbl:ablation-osfpa-without}} and Table~{\ref{tbl:ablation-osfpa-with}}. Several observations can be made. First, comparing Id1 and Id2, finetuning with perception-domain loss consistently improves all objective metrics over that in standard FM-based one-step generation. This is reasonable as standard FM optimizes predictions across all timesteps and OS-FPA directly focuses on the beginning timestep. Second, perception-only finetuning can substantially improves non-intrusive scores, but may introduce noticeable spectral artifacts, as illustrated in Fig.~{\ref{fig:one-step-spec}}. Incorporating additional fidelity-domain constraints can alleviate this issue, although at the cost of modest performance degradations in perceptual quality (\emph{e.g.}, Id2 vs. Id4). Meanwhile, only adopting fidelity-domain loss can improve performance but is generally inferior to the dual-domain setting. These results demonstrate the significance of jointly considering reconstruction fidelity and perceptual quality. We also compare between different pretrained SSL models for perception-domain alignment, including WavLM-Large, Wav2vec 2.0-Large, and HuBERT-Large. As indicated in Id4-Id6, comparable performance is achieved across three cases without a consistently superior choice. As a result, we empirically adopt WavLM-Large as the default configuration.
\vspace{-0.5cm}
\subsection{Objective Comparisons for Speech Denoising and Dereverberation Tasks}\label{sec:comparisons-baselines}
Based on the analysis above, the fine-tuned single-step counterpart of the BaME is adopted by default. In this part, we first compare the performance on denoising and dereverberation tasks. In Table~{\ref{tbl:objective_comparisons-dns-wo-reverb}}, we present objective performance on DNS-Challenge 2020 \textit{without reverb} test set. Several observations can be made. First, discriminative methods remain overall competitive in intrusive metrics, while advanced generative methods generally achieve better perceptual scores. For example, BSRNN achieves the highest eSTOI, whereas Metis obtains the highest scores in several non-intrusive metrics. Second, generative methods operating on continuous representations, such as Mel and SSL, generally exhibit better fidelity quality over discrete-token-based methods. Besides, we observe that the PASE family (including PASE, StuPASE, PhaseFlow and UniPASE) achieves relatively lower WER, demonstrating the benefit of auxiliary semantic information. Notably, the proposed best results on five intrusive metrics, with only eSTOI slightly inferior to BSRNN. Despite somewhat lower perceptual quality scores than Metis, our proposed BaME maintains competitive performance, indicating a favorable fidelity-perception trade-off. 

In Table~{\ref{tbl:objective_comparisons-dns-wi-reverb}}, we present the performance on \textit{with reverb} test set. Similarly, BaME exhibits the best scores in LPS, SBS, and SpkSim, and is slightly inferior to CleanMel in PESQ, eSTOI and WER. The reduced advantage may be partly attributed to the unreliable degraded phase under reverberation. Nonetheless, BaME still exhibits competitive scores in multiple non-intrusive metrics, demonstrating the effectiveness under complicated acoustic degradations. 

Fig.~{\ref{fig:sota_spectra_comparisons}} presents spectral visualizations of different methods. As observed, the discriminative method BSRNN introduces severe spectral distortions while generative methods such as LLaSE-G1 and AnyEnhance also exhibit missing spectral components in the mid- and high-frequency regions, as indicated in the blue boxes. Although more advanced generative approaches like Metis and UniSE exhibit promising perceptual quality, they may introduce hallucinated speech content and alter the original acoustic characteristics, as indicated by the black boxes. In contrast, the proposed BaME better preserves the target spectral structures while maintaining perceptually plausible details, demonstrating a favorable balance between fidelity and perceptual quality. Note that despite severe acoustic degradation, the degraded speech exhibits a relatively low WER, which is possibly due to the training procedure of Whisper-Large model on diverse audio data. Nonetheless, generative speech hallucinations may alter linguistic content and consequently lead to higher WER.

Table~{\ref{tbl:objective_comparisons_ears}} presents the results on EARS-WHAM\_v2 and EARS-Reverb\_v2 test sets, corresponding to denoising and dereverberation tasks, respectively. Note that only a set of the baselines is listed due to the lack of pretrained model weights. Notably, the proposed method yields the best PESQ and eSTOI on both benchmarks, demonstrating its effectiveness in preserving acoustic fidelity under different degradation conditions. Meanwhile, it maintains competitive non-intrusive scores and is slightly inferior to Metis. 
\renewcommand\arraystretch{1.20}
\begin{table*}[t]
	\centering
	\Huge
	\caption{Comparison of model complexity and inference efficiency. Inference time is measured in seconds, with the corresponding real-time factor (RTF). ``a$\times$'' denotes the speed-up ratio over real time. The RTF of CleanMel on CPU is not provided as it adopts Mamba block, which relies on optimized CUDA kernels and does not provide a directly comparable CPU inference configuration.}
	\label{tbl:complexity_comparison}
	\resizebox{\textwidth}{!}{
		\begin{tabular}{lccccccccccccccc}
			\toprule
			\textbf{Models}
			& TF-GridNet & BSRNN & CleanMel & StoRM & VoiceFixer
			& FlowSE & SenSE & GenSE & LLaSE-G1 & Metis
			& PASE & UniPASE & StuPASE & UniSE & \textbf{BaME} \\
			\midrule
			\textbf{\#Params (M)}
			& 8.52 & 63.1 & 20.4 & 55.1 & 122.1
			& 350.6 & 1575.7 & 667.6 & 1895.6 & 2134.2
			& 382.1 & 543.7 & 565.6 & 263.9 & 332.4 \\
			\textbf{MACs (Giga/s)}
			& 134.7 & 60.6 & 65.5 & 31539.0 & 61.8
			& 1292.5 & 475.3 & 99.0 & 63.9 & 1660.0
			& 16.9 & 25.0 & 107.2 & 42.2 & 45.5 \\
			\midrule
			\textbf{RTF (CPU)}
			& \shortstack{2.95\\(0.34$\times$)}
			& \shortstack{1.48\\(0.68$\times$)}
			& -
			& \shortstack{677.04\\(0.001$\times$)}
			& \shortstack{0.49\\(2.03$\times$)}
			& \shortstack{19.12\\(0.05$\times$)}
			& \shortstack{16.31\\(0.06$\times$)}
			& \shortstack{321.35\\(0.003$\times$)}
			& \shortstack{0.94\\(1.06$\times$)}
			& \shortstack{7.24\\(0.24$\times$)}
			& \shortstack{0.39\\(2.52$\times$)}
			& \shortstack{0.42\\(2.36$\times$)}
			& \shortstack{1.82\\(0.55$\times$)}
			& \shortstack{2.96\\(0.28$\times$)}
			& \shortstack{0.85\\(1.18$\times$)} \\
			\textbf{RTF (GPU)}
			& \shortstack{0.061\\(16.41$\times$)}
			& \shortstack{0.031\\(31.85$\times$)}
			& \shortstack{0.029\\(33.94$\times$)}
			& \shortstack{3.624\\(0.28$\times$)}
			& \shortstack{0.012\\(85.8$\times$)}
			& \shortstack{0.566\\(1.77$\times$)}
			& \shortstack{0.56\\(1.79$\times$)}
			& \shortstack{3.44\\(0.29$\times$)}
			& \shortstack{0.013\\(78.88$\times$)}
			& \shortstack{0.36\\(2.75$\times$)}
			& \shortstack{0.012\\(80.39$\times$)}
			& \shortstack{0.007\\(152.28$\times$)}
			& \shortstack{0.092\\(10.82$\times$)}
			& \shortstack{0.61\\(1.64$\times$)}
			& \shortstack{0.025\\(40.78$\times$)} \\
			\bottomrule
		\end{tabular}
	}
	\vspace{-0.3cm}
\end{table*}

\vspace{-0.25cm}
\subsection{Objective Comparisons for Universal Speech Enhancement Tasks}\label{sec:objective-universal}
In this part, we further investigate the performance on the universal SE task, \emph{i.e.}, various degradations are involved. In Table~{\ref{tbl:objective_comparisons-gsr}}, we show the results on the GSR test set, which involve diverse acoustic degradations. Note that we select only six baselines as other models either do not support universal SE or lack publicly available pretrained weights. Again, the proposed BaME yields the best results on all intrusive metrics and achieves suboptimal performance on non-intrusive metrics and is slightly inferior to Metis. Overall, these results demonstrate the effectiveness of the proposed method in universal acoustic degradation scenarios.    

Besides, we also compare with the top tiers in the URGENT Challenge, one of the most representative challenges for universal SE task in more recent years. As shown in Table~{\ref{tbl:objective_comparisons-urgent}}, compared with baselines, the proposed BaME achieves the best or second-best performance across all reported metrics, validating its effectiveness under diverse acoustic degradations. Notably, the top three systems predominantly adopt discriminative or hybrid discriminative-generative paradigms, which indicate that guaranteeing reliable speech restoration under complex acoustic scenarios remains rather challenging for pure generative SE methods.

Overall, BaME outperforms the top-ranked challenge systems in non-intrusive metrics. However, a performance gap remains in several intrusive metrics, and further optimization is still needed.
\begin{figure}[t]
	\centering
	\includegraphics[width=0.95\columnwidth]{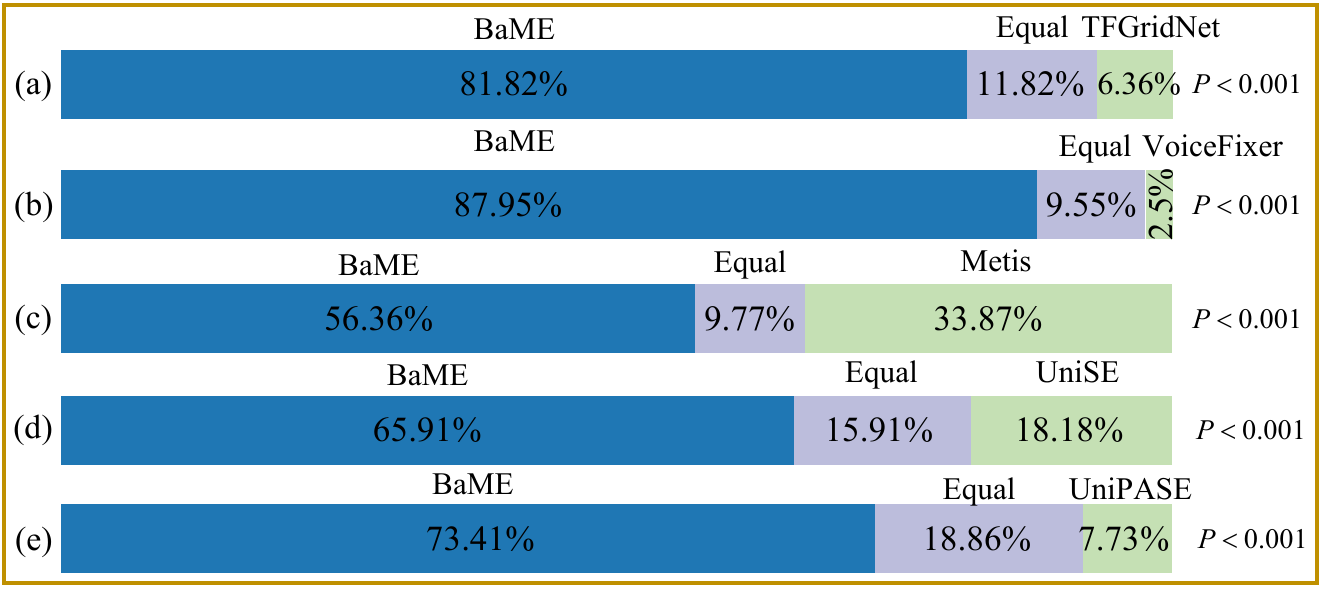}
	\vspace{-0.25cm}
	\caption{Absolute preference test results between the proposed BaME and five representative baselines.
	}
	\label{fig:absolute_mos_abtest}
	\vspace{-0.28cm}
\end{figure}
\begin{figure}[t]
	\centering
	\includegraphics[width=0.95\columnwidth]{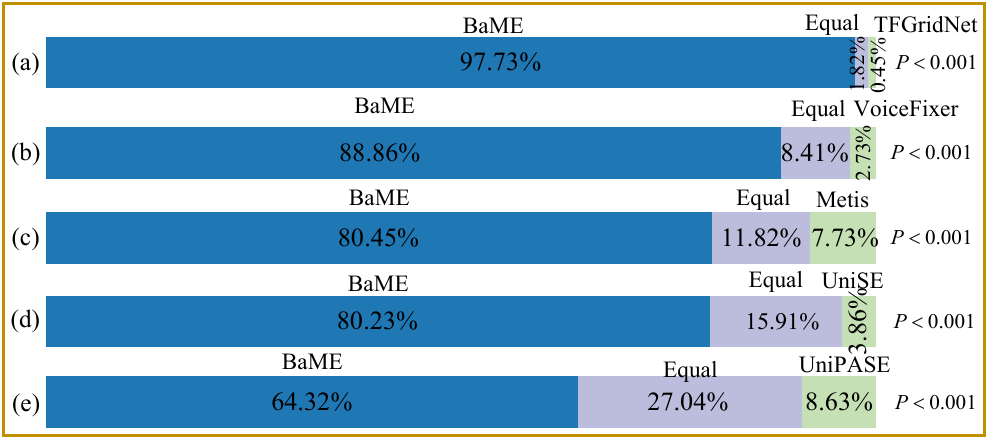}
	\vspace{-0.25cm}
	\caption{Similarity preference test results between the proposed BaME and five representative baselines.
	}
	\label{fig:similar_mos_abtest}
	\vspace{-0.35cm}
\end{figure}
\vspace{-0.25cm} 
\subsection{Subjective Comparisons with Advanced Baselines}\label{sec:subjective-evaluation}
In addition to objective evaluations, we conduct two types of subjective A/B preference tests. To segregate subjective quality in terms of perceptual quality and fidelity, we design two types of A/B preference tests, namely Absolute Preference Test and Similarity Preference Test. The former measures overall listening quality without providing clean references, whereas the latter requires listeners to select the item that is more similar to the provided reference. For detailed testing procedure, please refer to Sec.~{\ref{sec:metrics}}.

Fig.~{\ref{fig:absolute_mos_abtest}} and Fig.~{\ref{fig:similar_mos_abtest}} show the preference results of two scenarios, respectively. For absolute subjective quality, BaME is consistently preferred over five representative baselines with statistical significance, \emph{i.e.}, $p<0.001$. Note that although Metis achieves relatively better performance on several non-intrusive metrics, most listeners still favor BaME, indicating that objective metrics may not fully align with human ratings. For similarity subjective quality, BaME again exhibits clear superiority over all five baselines, with particularly pronounced preference margins against Metis and UniPASE. These results further reveal the superiority of BaME in both perceptual quality and fidelity. For interested readers, we provide collected subjective results and subjective evaluation platform in \textcolor{blue}{https://huggingface.co/AndongLi/BaME/tree/main}.
\subsection{Model Size and Computational Complexity}\label{sec:model-size-computational-complexity}
Table~{\ref{tbl:complexity_comparison}} compares the model parameters, computational overhead, and inference efficiency of different approaches in terms of real-time factor (RTF). The CPU and GPU inference times are measured on an Intel Xeon Platinum 8336C CPU (2.30 GHz) and an NVIDIA A100 GPU, respectively. As observed, BaME yields moderate computational complexity despite its relatively large parameters, which mainly stems from the DeWavLM semantic extractor. As illustrated in Sec.~{\ref{sec:effect-semantic}}, replacing DeWavLM with a lightweight semantic extractor can substantially reduce the model size, albeit with some performance degradation. Further, due to the one-step generation, BaME enjoys real-time inference on CPU with an RTF of 0.85 (1.18$\times$ speedup), outperforming advanced discriminative methods such as TF-GridNet and BSRNN in inference efficiency. It achieves a processing speed of 40.78$\times$ real time on GPU, demonstrating its practical efficiency.
\begin{figure}[t]
	\centering
	\includegraphics[width=\columnwidth]{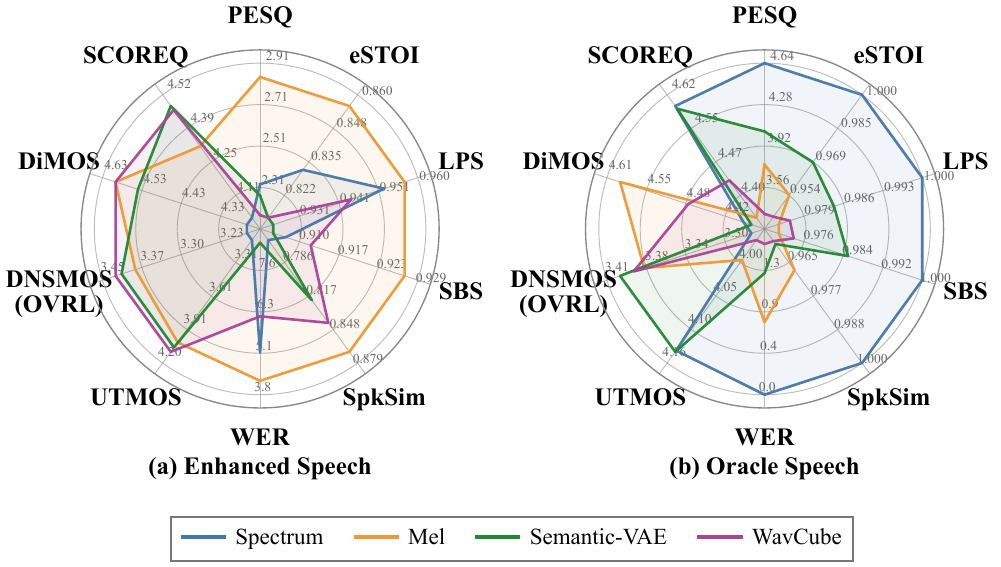}
	\vspace{-0.25cm}
	\caption{Comparisons of Mel with three alternative input representations: complex-valued spectrum, VAE latent features from Semantic-VAE, and latent features from WavCube. (a) Enhancement performance across ten objective metrics, averaged over the DNS-Challenge \textit{without reverb} and \textit{with reverb} test sets. (b) Oracle reconstruction performance using the corresponding ground-truth representations as inputs. Note that the WER axis is reversed so that larger radii consistently indicate better performance across all metrics. The two radar charts are independently normalized for visualization.
	}
	\label{fig:radar}
	\vspace{-0.35cm}
\end{figure}
\begin{table}[t]
	\centering
	\Huge
	\caption{Comparison of parameters and computational complexity for encoder and decoder across different representation types. ``-'' denotes no learnable encoder or decoder is adopted.}
	\label{tbl:representation_complexity}
	\resizebox{0.95\columnwidth}{!}{
		\begin{tabular}{lcccc}
			\toprule
			\textbf{Representation}
			& \multicolumn{2}{c}{\textbf{Encoder}}
			& \multicolumn{2}{c}{\textbf{Decoder}} \\
			\cmidrule(lr){2-3}\cmidrule(lr){4-5}
			\textbf{Type} & \textbf{\#Params(M)} & \textbf{MACs(Giga/s)}
			& \textbf{\#Params(M)} & \textbf{MACs(Giga/s)} \\
			\midrule
			Complex Spectrum     &- &- &- &- \\
			Mel          &- &- &\textbf{5.8} &\textbf{5.55} \\
			Semantic-VAE &21.6 &7.24 &50.6 &23.30 \\
			WavCube       &240.6 &20.54 &315.6 &17.35 \\
			\bottomrule
		\end{tabular}
	}
	\vspace{-0.25cm}
\end{table}

\vspace{-0.25cm}
\subsection{Comparisons on Different Input Representations}\label{sec:comparisons-input}
To further investigate the effectiveness of Mel representation in balancing fidelity and perceptual quality, we compare it with another three alternative representations: complex-valued spectrum, VAE-based continuous latents, and SSL-derived continuous latents. For the latter two, we utilize Semantic-VAE~{\cite{niu26_interspeech}} and WavCube~{\cite{yang2026wavcube}}, respectively. Semantic-VAE operates at 16 kHz with a frame rate of 40 Hz and a latent dimension of 64 and that of \{16 kHz, 50 Hz, 128\} for WavCube. For complex-valued spectrum, we adopt the non-uniform subband division strategy in~{\cite{li2025bridgevoc}} to construct 25 subbands, followed by a subband-aware generative backbone similar to SaT. For the two learned latent representations, which lack an explicit frequency-band organization, we adopt a DiT backbone similar to~{\cite{wang25s_interspeech}}. All cases incorporate DeWavLM for semantic conditioning and follow the same training configurations as BaME. $\text{NFE}=8$ is adopted during inference by default. 

Fig.~{\ref{fig:radar}}(a) shows the enhancement performance across different representations. Notably, Mel achieves the best performance on the intrusive metrics while yielding suboptimal scores on the non-intrusive metrics, indicating a favorable fidelity-perception trade-off. Despite the lossless nature of complex-valued spectrum, it exhibits considerably worse performance than Mel-domain modeling. It can be attributed to the higher dimensionality of complex spectrum, which heavily increases the difficulty of generative modeling, as well as the strong reconstruction prior by the pretrained neural vocoder.  

To further investigate whether these differences arise from the reconstruction capability of the representations themselves, we evaluate oracle reconstruction by directly decoding clean representations, as shown in Fig.~{\ref{fig:radar}}(b). As observed, complex spectrum allows nearly perfect reconstruction, whereas the compressed representations exhibit varying degrees of reconstruction loss. Nonetheless, the differences in oracle reconstruction are generally less pronounced than those observed in enhanced speech, indicating that reconstruction capability alone cannot fully explain the enhancement performance. Rather, the suitability of a representation for generative modeling also plays an important role.

Table~{\ref{tbl:representation_complexity}} shows the parameters and computational complexity of the encoders and decoders across different representations. Unlike learned latent representations, Mel requires no learnable encoder and employs a lightweight neural vocoder, substantially reducing representation-related computational overhead. Although complex spectra also require no learnable encoder or decoder, their high dimensionality increases the difficulty of generative modeling. These results further highlight the potential of Mel representations for efficient generative SE, particularly in resource-constrained scenarios.
\section{Conclusion}\label{sec:conclusion}
In this paper, we revisit Mel spectrum as a structured acoustic representation for generative speech enhancement and develop BaME to exploit their inherent frequency-band structure and complementary information from degraded observations. First, we leverage the explicit frequency-band distribution of Mel representation and propose a lightweight SaT backbone that achieves superior performance over a substantially larger DiT backbone. we incorporate degraded-phase information into a phase-aware neural vocoder and propose a quality-guided dual-phase vocoding strategy to mitigate the adverse effects of unreliable phase priors under complicated acoustic conditions. Third, we develop a simple yet effective one-step generation strategy that jointly optimizes signal-level fidelity and perceptual quality. Extensive objective and subjective evaluations across multiple benchmarks demonstrate the effectiveness of BaME in balancing acoustic fidelity and perceptual quality. These findings suggest that learnable latent representations based on complex neural audio tokenzier are not necessarily prerequisites for competitive generative speech enhancement, particularly when acoustic fidelity and perceptual quality are joinly considered. Instead, explicitly exploiting the inherent structure of conventional Mel representations provides a promising alternative for faithful and efficient speech restoration.

\textbf{Limitations}: Existing objective metrics do not yet provide a ``golden'' measure that accurately reflects acoustic fidelity and perceptual quality, motivating further research into more accurate and comprehensive evaluation criteria. Besides, our findings do not advocate the optimality of Mel representation for generative speech enhancement, and it remains an significant direction to identify better and more suitable acoustic representations for speech front-end tasks.
\vspace{-0.25cm}
\section{Acknowledgement}\label{sec:acknow}
The authors would like to thank Xiaobin Rong, Zheng Wang, and Kelan Kuang for kindly providing the enhanced speech samples generated by PhASE-Flow, L3-SE and AnyEnhance, and Xingchen-UniFE, respectively, on the DNS-Challenge test sets.


\bibliographystyle{IEEEtran}
\bibliography{refs}
\vspace{-10mm}

\begin{IEEEbiography}[{\includegraphics[width=0.95in,height=1.25in,clip,keepaspectratio]{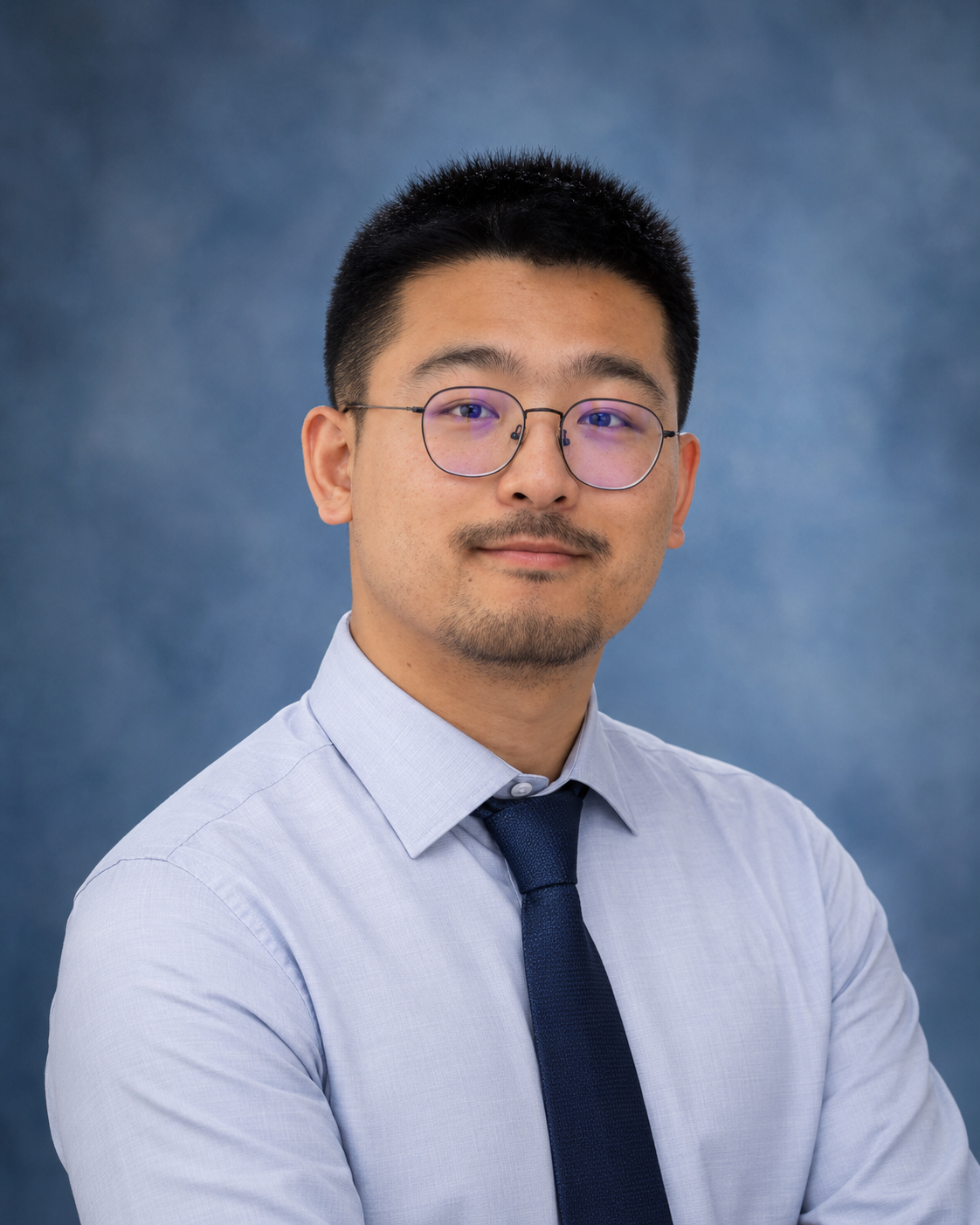}}]
	{Andong Li} 
	(Member, IEEE) received the B.S. degree in information engineering from Southeast University, Nanjing, China, in 2018, and the Ph.D. degree in signal and information processing from the Institute of Acoustics, Chinese Academy of Sciences, Beijing, China, in 2023. From 2023 to 2024, he was a Senior Researcher with Tencent AI Lab. He is currently an Assistant Professor with the Institute of Acoustics, Chinese Academy of Sciences. His research interests include speech enhancement, audio coding, array signal processing, and speech editing. He is also an active reviewer for multiple leading conferences and journals, such as INTERSPEECH, ICASSP, ACM MM,
	SLT, ASRU, IEEE Signal Processing Letters, Speech Communication, Neural Networks, and IEEE/ACM Transactions on Audio, Speech, and Language Processing.
\end{IEEEbiography}
\vspace{-10mm}
\begin{IEEEbiography}[{\includegraphics[width=0.95in,height=1.25in,clip,keepaspectratio]{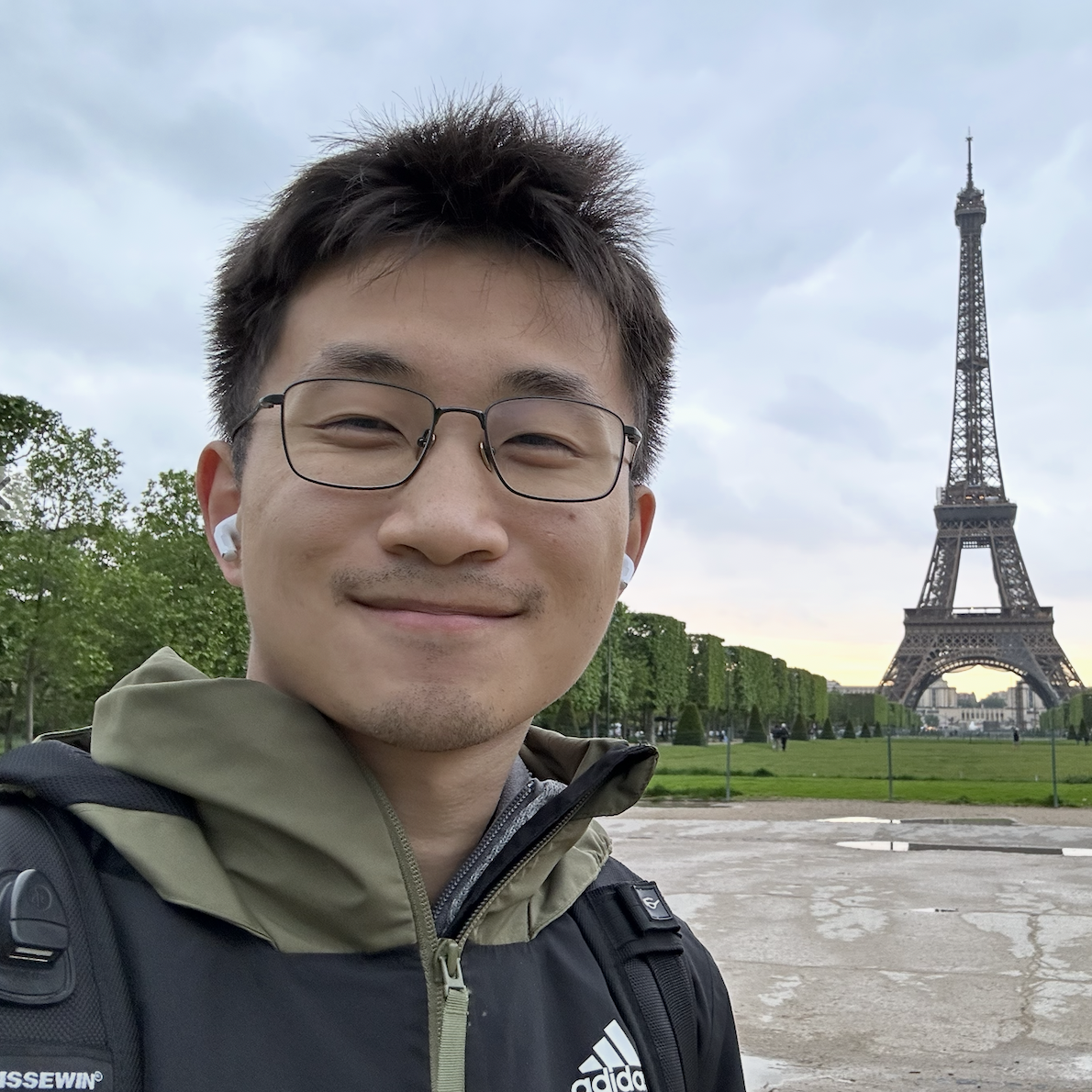}}]
	{Kun Zhou}
	(Member, IEEE) received the B.Eng. degree from the University of Electronic Science and Technology of China (UESTC) in 2018, and the M.Sc. and Ph.D. degrees from the National University of Singapore (NUS) in 2019 and 2023, respectively. He is currently a Postdoctoral Researcher at Télécom Paris, France, affiliated with Hi! PARIS. Previously, he was a Research Engineer at Tongyi Lab, Alibaba Group (2023–2026), and a Visiting Ph.D. Student at the University of Bremen (2023) and the University of Texas at Dallas (2022). His research focuses on speech generation. He received the PREMIA Best Student Paper Award in 2022 and the Best Overview Paper Award from Speech Communication. He has also served as an organizing committee member and reviewer for leading international conferences and journals.
\end{IEEEbiography}
\vspace{-10mm}
\begin{IEEEbiography}[{\includegraphics[width=0.95in,height=1.25in,clip,keepaspectratio]{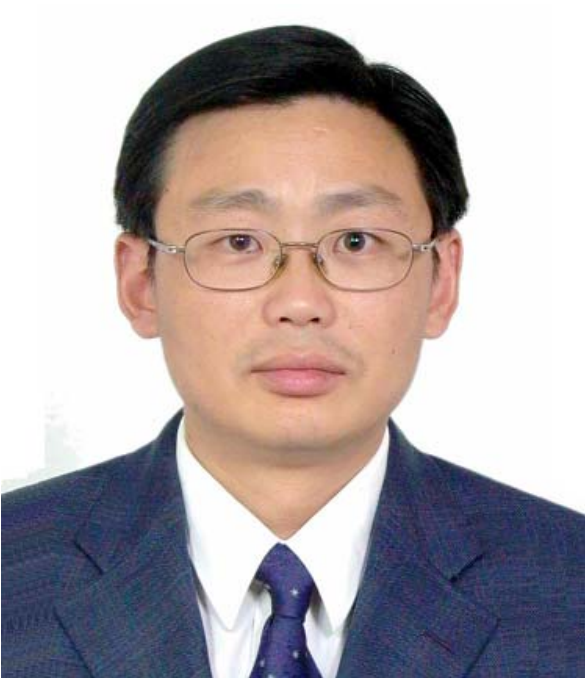}}]
	{Xiaodong Li}
	received the B.S. degree from Nanjing University, Nanjing, China, in 1988, the M.Eng. degree from the Harbin Engineering University, Harbin, China, in 1991, and the Ph.D. degree in physical acoustics from the Institute of Acoustics, Chinese Academy of Sciences (IACAS), Beijing, China, in 1995. After a short period as a Research Fellow with the Hong Kong Polytechnic University, Hong Kong, working on active control of sound, he was appointed Assistant Professor with IACAS in 1997. He was made an Associate Professor with IACAS in 1998 and Professor in 2002. His research interests include acoustic signal processing, active control of sound and vibration, and engineering acoustics.
\end{IEEEbiography}
\vspace{-10mm}
\begin{IEEEbiography}[{\includegraphics[width=0.95in,height=1.25in,clip,keepaspectratio]{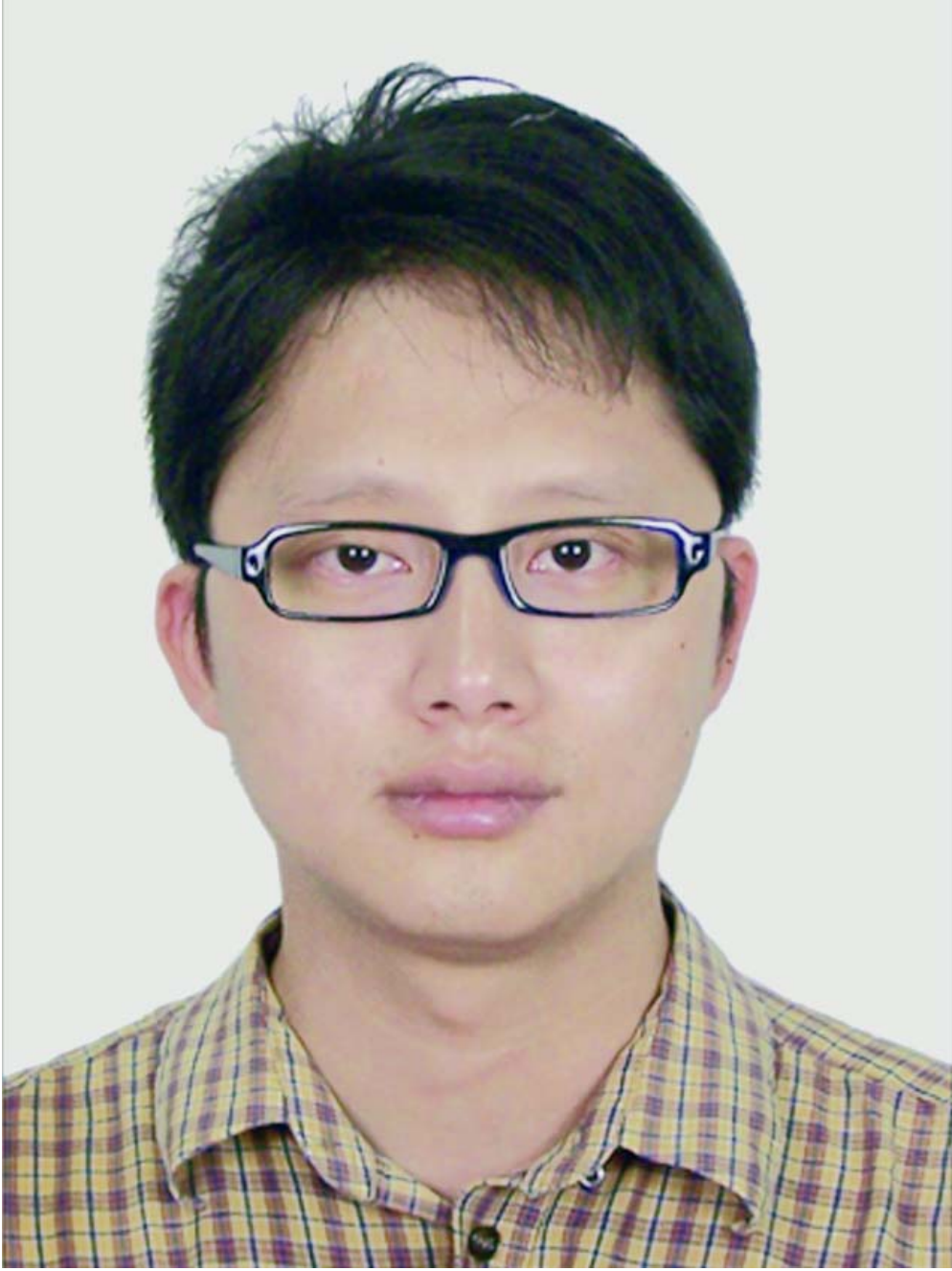}}]
	{Chengshi Zheng}
	(Senior Member, IEEE)
	received the B.S. degree in electronic engineering and information science from the University of Science and Technology of China, Hefei, China, in 2004, and the Ph.D. degree in acoustics from the Institute of Acoustics, Chinese Academy of Sciences, Beijing, China, in 2009. Since then, he has been with the Institute of Acoustics, Chinese Academy of Sciences, where he is currently a Full Professor with the Key Laboratory of Noise and Vibration Research. From 2014 to 2015, he was a Visiting Scientist with the Chair of Multimedia Communications and Signal Processing, University Erlangen-Nuremberg, Erlangen, Germany. His research interests include speech enhancement, deep learning, array signal processing, and signal processing for hearing aids. He was the recipient of the Outstanding Reviewer Awards at INTERSPEECH 2023.
\end{IEEEbiography}

\end{document}